\documentclass[journal,comsoc]{IEEEtran}
\usepackage{amssymb,amsmath,graphicx}
\usepackage{xcolor}

\usepackage{stmaryrd}

\usepackage[T1]{fontenc}
\usepackage{tipa}
\usepackage{subfig}
\usepackage{epstopdf}
\usepackage{booktabs} 
\usepackage{multirow}
\usepackage{algorithm}
\usepackage[noend]{algpseudocode}

\newcommand{\RevisionText}[1]{{#1}}

\title{Post-processing speech recordings during MRI}

\author{\IEEEauthorblockN{Juha Kuortti\IEEEauthorrefmark{1} and Jarmo Malinen\IEEEauthorrefmark{1}\IEEEauthorrefmark{2} and Antti Ojalammi\IEEEauthorrefmark{1}}\\
\IEEEauthorblockA{\IEEEauthorrefmark{1}School of Science, Department of Mathematics and Systems Analysis, Aalto University}\\
\IEEEauthorblockA{\IEEEauthorrefmark{2}School of Electrical Engineering, Department of Signal Processing and Acoustics, Aalto University}
\thanks {Manuscript received XX.XX.XX 
  Corresponding author: J.~Malinen (email:jarmo.malinen@aalto.fi}}

\makeatletter
\def\name#1{\gdef\@name{#1\\}}
\makeatother
\name{{\em Juha Kuortti, Jarmo Malinen, Antti Ojalammi}}

\begin{document}

\maketitle

\bibliographystyle{IEEEtran}
\begin{abstract}
  We discuss post-processing of speech that has been recorded during
  Magnetic Resonance Imaging (MRI) of the vocal tract area. These
  speech recordings are contaminated by high levels of acoustic noise
  from the MRI scanner.  Also, the frequency response of the sound
  signal path is not flat as a result of restrictions on recording
  instrumentation and arrangements due to MRI technology.  The
  post-processing algorithm for noise reduction is based on adaptive
  spectral filtering, and it has been designed keeping in mind the
  requirements of subsequent formant extraction.

  Speech material was used for validation of the post-processing
  algorithm, consisting of samples of prolonged vowel productions
  during the MRI.  The comparison data was recorded in the anechoic
  chamber from the same test subject. Spectral envelopes and formants
  were computed for the post-processed speech and the comparison
  data. Artificially noise-contaminated vowel samples (with a known
  formant structure) were used for validation experiments to determine
  performance of the algorithm where using true data would be
  difficult.  Resonances computed by an acoustic model and, similarly,
  those measured from 3D printed vocal tract physical models were used
  as comparison data as well.

  The properties of recording instrumentation or the post-processing
  algorithm do not explain the observed frequency dependent
  discrepancy between formant data from experiments during MRI and in
  the anechoic chamber. It is shown that the discrepancy is
  statistically significant, in particular, where it is largest at
  around $1 \, \mathrm{kHz}$ and $2 \, \mathrm{kHz}$. In order to
  evaluate the role of the reflecting surfaces of the MRI head coil,
  eigenvalues of the Helmholtz equation were solved by Finite Element
  Method in all vowel configurations of the vocal tract, using a
  digital head model and an idealised MRI coil model for the exterior
  space.  The eigenvalues corresponding to strong excitations of the
  exterior space were found to coincide with ``exterior formants''
  observed in speech recordings during the MRI scan. However, the role
  of test subject's adaptation to noise and constrained space
  acoustics during an MRI examination cannot be ruled out.
\end{abstract}

\section{Introduction}
\label{IntroSec}

Modern medical imaging technologies such as Ultrasono\-graphy (USG),
X-ray Computer Tomography (CT), and Magnetic Resonance Imaging (MRI)
have revolutionised studies of speech and articulation. There are,
however, significant differences in applicability and image quality
between these technologies. Considering the imaging of the whole
speech apparatus, the use of inherently low-resolution USG is often
impractical, and the high-resolution CT exposes the test subject to
potentially significant doses of ionising radiation.  MRI remains an
attractive approach for large scale articulation studies but there
are, unfortunately, many other restrictions on what can be done during
an MRI scan as discussed in
\cite{A-H-M-P-P-S-V:RSSAMRI,A-A-H-J-K-K-L-M-M-P-S-V:LSDASMRIS}.

Since the intra-subject variability of speech often appears to be of
the same magnitude as the inter-subject variability, it is desirable
to sample speech simultaneously with the MRI experiment in order to
obtain \emph{paired data}. Such paired data is a particularly valuable
asset in developing and validating a computational model for speech
such as proposed in \cite{A-A-M-M-V:MLBVFVTOPI}.  Unfortunately,
speech signal recorded during MRI contains many artefacts that are
mainly due to high acoustic noise level inside the MRI scanner. There
are additional artefacts due to the non-flat frequency response of the
MRI-proof audio measurement system and further challenges related to
the constrained space acoustics inside the MRI head and neck coils.

Noise cancellation is a classical subject matter in signal processing
that in the context of speech enhancement can be divided into two main
classes: \emph{adaptive noise cancellation} techniques and the
\emph{blind source separation} methods such as FastICA introduced in
\cite{H-O:ICAAA}.  The purpose of this article is to introduce,
analyse, and validate a post-processing algorithm of the former type
for treating speech that has been recorded during MRI.\footnote{Some
  experiments on the same speech data have been carried out using
  FastICA as well but adaptive methods seem to give better results.}
Compared to blind source separation, the tractability of the
processing algorithm favours adaptive noise cancellation that may take
place in time domain, in frequency domain, or partly in both.  The
algorithm discussed in this article is designed based on lessons
learned from an earlier algorithm introduced in
\cite[Section~4]{A-A-H-J-K-K-L-M-M-P-S-V:LSDASMRIS}.  For different
approaches for dealing with the MRI noise, see also
\cite{bresch,P-H-H:TMMNRRSDPMRID,P-P-F:ASPANPDMRI,I-B-I:TUNSMRIS} that
will be discussed at the end of the article.

\begin{figure}[t]
\begin{center}
  \includegraphics[width=87mm,height=50mm]{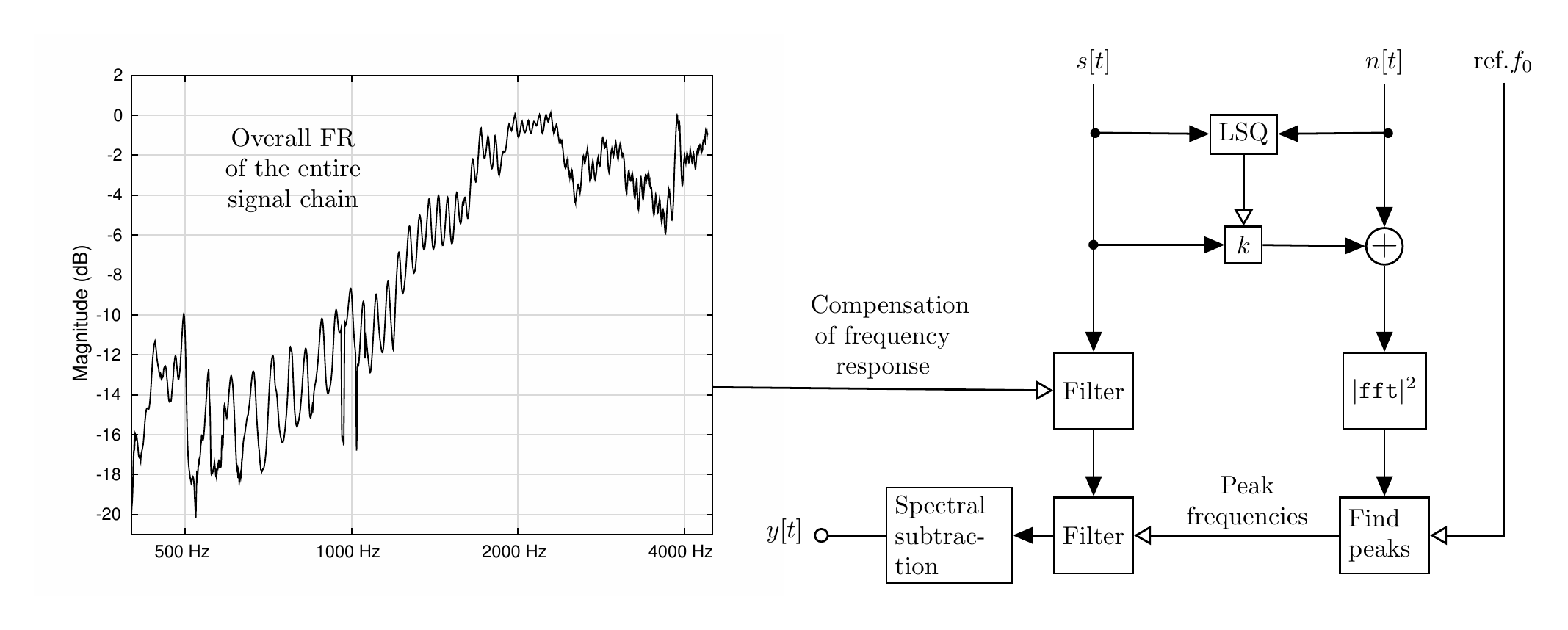}
\end{center}
\begin{center}
  \includegraphics[width=0.23\textwidth]{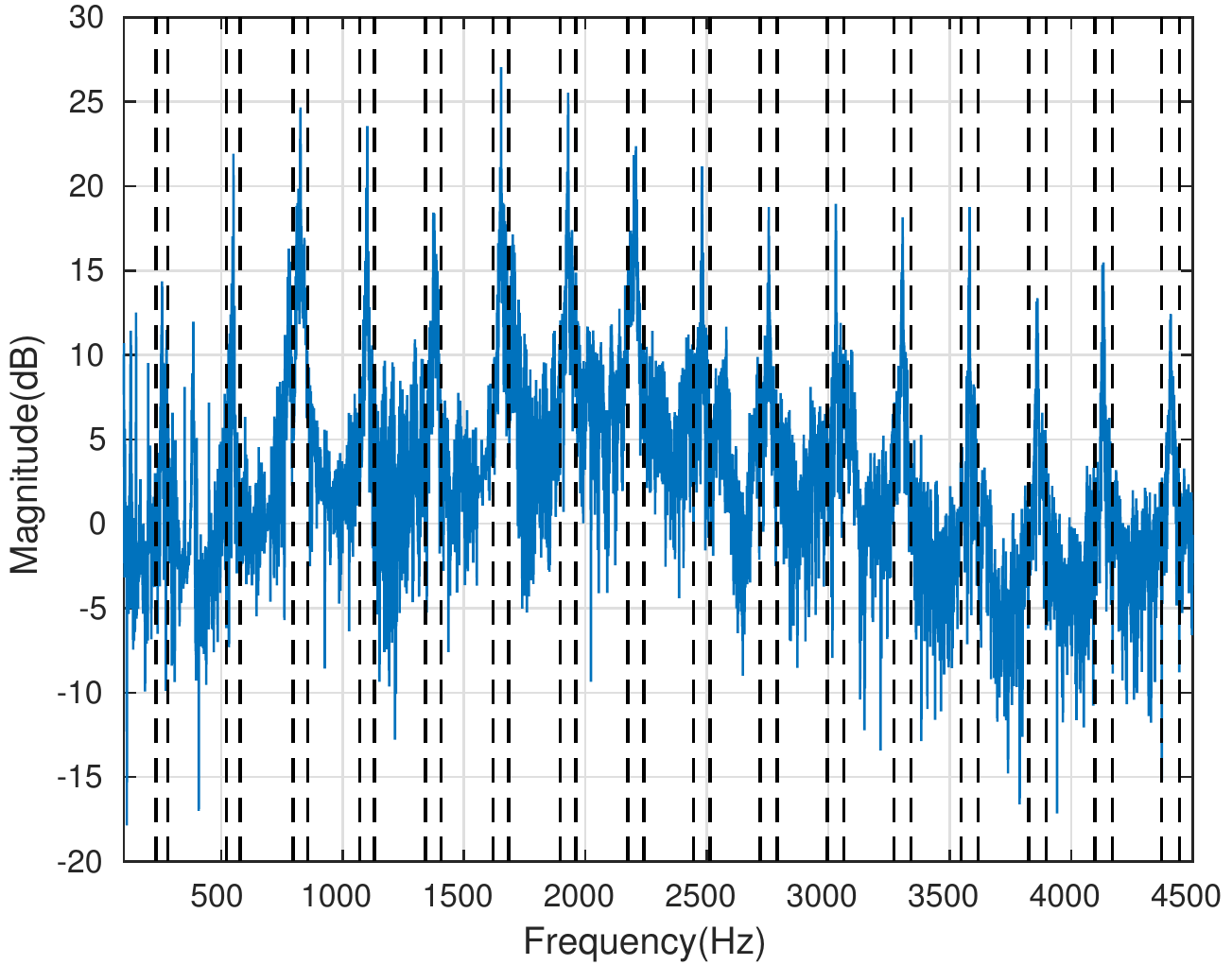}
 \includegraphics[width=0.22\textwidth]{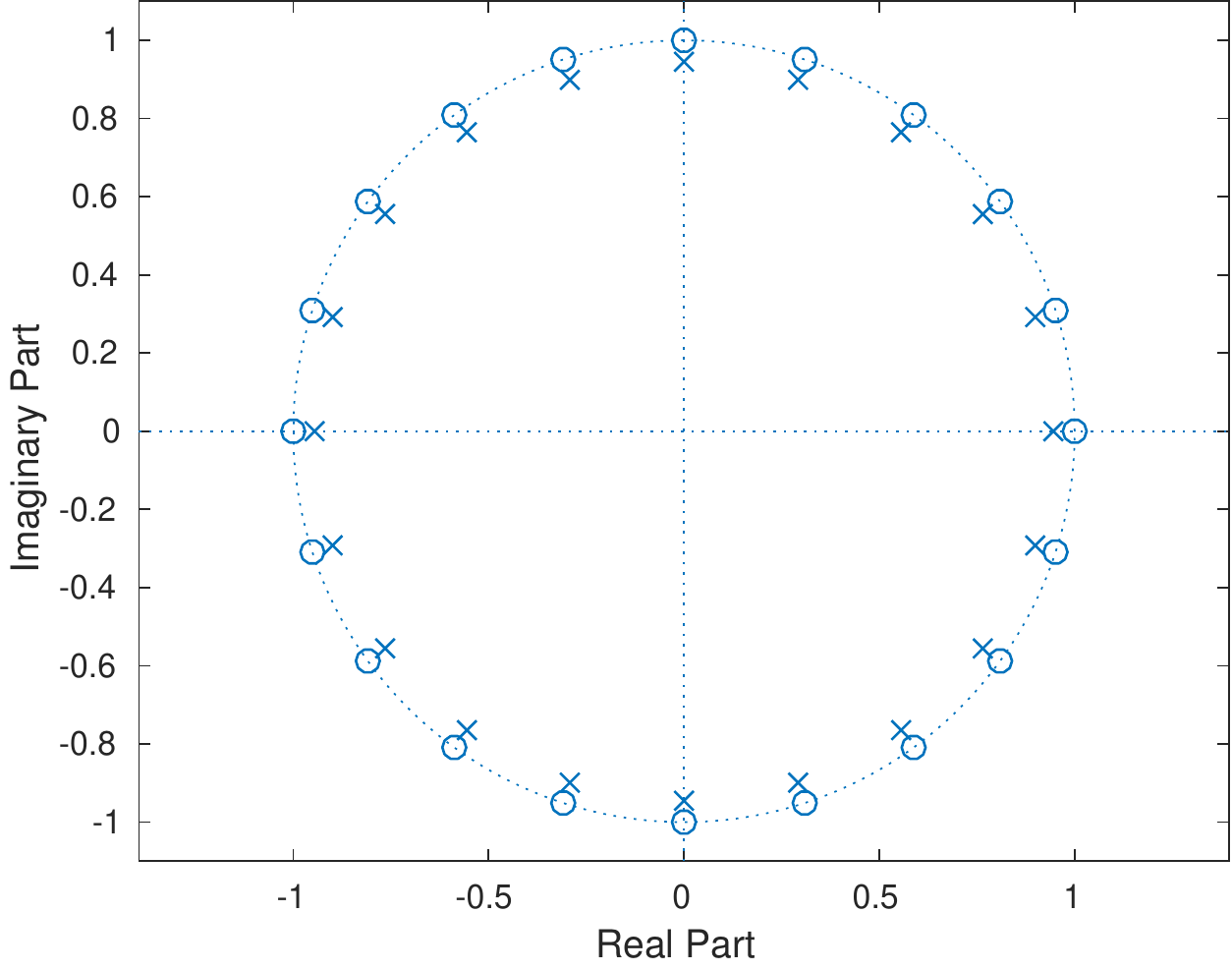}
\end{center}
\caption{\label{AlgorithmNoiseFig} Upper panel: A block diagram of the
  post-processing algorithm. Here $s[t]$ and $n[t]$ denote the
  discretised speech and noise samples at
  $f_s = 44\, 100 \, \textrm{Hz}$, respectively. The signal $y[t]$ is
  de-noised speech.  Lower panel on the left: Harmonic structure of
  the MRI noise and stop bands estimated from it. Lower panel on the
  right: The zero/pole placement in $z$-plane of the notch filter of
  degree $20$ for removing the frequency $f_s/20$ and its harmonics
  below the Nyquist frequency $f_s/2$.}
\end{figure}

When designing a practical solution, one should consider, at least,
these three aspects of the noise cancellation problem: \textrm{(i)}
what kind of noise should be rejected, \textrm{(ii)} what kind of
signal or signal characteristic should be preserved, and
\textrm{(iii)} how the resulting de-noised signal is to be used.  In
this work, the noise is generated by an MRI scanner, the preserved
signal consists of prolonged, static vowel utterances, and the
de-noised signals should be usable for high-resolution spectral
analysis of speech formants.  The noise spectrum of the MRI scanner
(in these experiments, Siemens Magnetom Avanto 1.5T) has a lot of
harmonic structure on few discrete frequencies as shown in
Fig.~\ref{AlgorithmNoiseFig} (lower panel), and it changes during the
course of the MRI scan. The proposed algorithm estimates the harmonics
of the noise, and removes their contribution by tight notch filters as
explained in Fig.~\ref{AlgorithmNoiseFig}.  There are additional
heuristics to prevent the removal of multiples of the fundamental
glottal frequency ($f_0$) of the speech that, unfortunately, somewhat
resemble the noise spectrum of the MRI scanner. One of the caveats is
not to have the algorithm ``bake'' noise energy into spurious spectral
energy concentrations that would skew the true formant content -- this
may be a serious cause of worry in non-linear signal processing that
is able to move energy from one frequency band to another.

Since the de-noised vowel data is used in, e.g.,
\cite{A-A-H-J-K-K-L-M-M-P-S-V:LSDASMRIS,K-K-M-O:MIOVTRV} for parameter
estimation and validation of a computational model, it is imperative
that the extracted formant positions, indeed, reflect precisely the
acoustic resonances of the corresponding MRI geometries of the vocal
tract. For model validation, the proposed post-processing algorithm is
applied to noisy speech data consisting of prolonged vowel samples
from which vowel formants should be extracted without bias.  In a
typical speech sample, the noise component is of a comparable level as
the speech component, but there is great variance between different
test subjects and even between different vowels from the same test
subject: A smaller mouth opening area results in lower emission of
sound power.


The outline of this article is as follows: After the data acquisition
has been described in Section~\ref{RecordingSec}, the post-processing
algorithm is described in Section~\ref{CancellationSec}.  The
validation of the algorithm is carried out in Section~\ref{PerfSec}
through four different approaches: \textrm{(i)} accuracy of the
formant extraction using a synthetic test signal with known formant
structure, \textrm{(ii)} comparison of spectral tilts (i.e., the
roll-off) of de-noised speech recorded during the MRI to similar data
recorded in the anechoic chamber, \textrm{(iii)} comparison of the
formants from de-noised speech to computationally obtained resonances
(see \cite{K-K-M-O:MIOVTRV}) as well as to spectral peaks measured
from 3D printed physical models from the simultaneously obtained MRI
geometries, and finally \textrm{(iv)} a perceptual vowel
classification experiment (see \cite{P-A-A-H-M-S-V:AFVRMRISD}) based
on de-noised speech recorded during the  MRI. These four validation
experiments support the conclusion that the proposed noise
cancellation algorithm can be used with good confidence for, at least,
obtaining formants from speech contaminated by MRI noise.
In Section~\ref{FormSubSec}, we apply the post-processing algorithm to
speech that has been recorded during MRI scans as detailed in
\cite{A-A-H-J-K-K-L-M-M-P-S-V:LSDASMRIS}. The objective is no longer
to validate the algorithm rather than to draw conclusions about the
speech data itself.  We again use comparison samples that have been
recorded in the anechoic chamber.  There is a statistically
significant ($p > 0.95$) discrepancy between some of the vowel
formants extracted from these two kinds of data. It is further
observed that the formant discrepancy has a consistent frequency
dependent behaviour shown in Fig.~\ref{VowelDiscrepancy} with steps at
around $1 \mathrm{kHz}$ and $2 \mathrm{kHz}$. In
Section~\ref{IdentificationSec}, a computational study is carried out
based on the Helmholtz equation and the exterior space model shown in
Fig.~\ref{exterior}. It is observed that the acoustic space between
the test subject's head and the MRI head coil produces a family of
spectral energy concentrations. They appear as a common feature (i.e.,
as ``external formants'') in vowel recordings during MRI but not in
similar recordings carried out in the anechoic chamber. In particular,
the frequencies $1 \mathrm{kHz}$ and $2 \mathrm{kHz}$ get identified
as external formants near some of the true vowel formants, explaining
the increased formant discrepancy observed in
Fig.~\ref{VowelDiscrepancy}.

\section{Speech recording during MR imaging}
\label{RecordingSec}

\subsection{Arrangements}
\label{ArrSubSec}

\begin{figure}[t]
\begin{center}
  \includegraphics[width=42mm,height=52mm ]{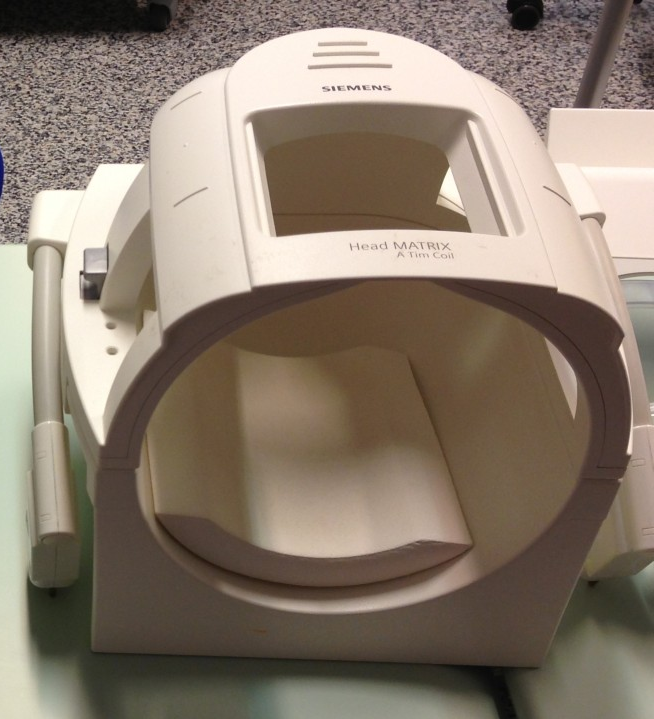}
  \vspace{0.2cm}
  \includegraphics[width=42mm,height=52mm]{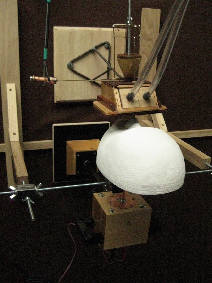}
\end{center}
\caption{\label{ArrangementDetail} Left panel: The MRI head coil of
  Siemens Magnetom Avanto 1.5T scanner.  The two-channel acoustic sound
  collector fits exactly the opening on the top. Right panel: The
  sound collector positioned above a head model similarly as in the
  MRI experiments.  The noise sample is acquired using a horn on the
  top surface of the collector and the speech sample from another
  similar horn pointing downwards.}
\end{figure}

The experimental arrangement has been detailed in
\cite{JP:WEMVMV,A-H-M-P-P-S-V:RSSAMRI,A-A-H-J-K-K-L-M-M-P-S-V:LSDASMRIS}.
Briefly, a two-channel acoustic sound collector samples speech and MRI
noise in a configuration shown in Fig.~\ref{ArrangementDetail}.  The
signals are acoustically transmitted to a microphone array inside a
sound-proof Faraday cage by wave\-guides of length $3.00$ m. The
microphone array contains electret microphones of type Panasonic
WM-62. The preamplification and A/D conversion of the signals is
carried out by conventional means, see
\cite[Section~3.1]{A-A-H-J-K-K-L-M-M-P-S-V:LSDASMRIS}. The experiments
were carried out using Siemens Magnetom Avanto 1.5T using 3D VIBE
(Volumetric Interpolated Breath-hold Examination) MRI sequence [58] as
it allows for sufficiently rapid static 3D acquisition.  Imaging
parameters, etc., have been described in
\cite[Section~3.2]{A-A-H-J-K-K-L-M-M-P-S-V:LSDASMRIS}.

\subsection{Phonetic and geometric materials}
\label{PhonSubSec}

The speech materials consist of Finnish vowels \textipa{[\textscripta
    , e, i, o, u, y, \ae , \oe]} that were pronounced by a 26-year-old
healthy male (in fact, the first author) in supine position during the 
MRI. The number of samples varies between 3 and 9 depending on the
vowel.  The MRI sequence requires up to $11.6$ s of continuous
articulation in a stationary supine position.  The test subject
produced the vowels at a fairly constant fundamental frequency $f_0$,
given by the cue signal to the earphones. Two different pitches $f_0 =
104 \, \mathrm{Hz}$ and $f_0 = 130 \, \mathrm{Hz}$ were used, and they
had been chosen so as to avoid spectral peaks of the MRI noise.

The paired MRI/speech data for this article was acquired during a
single session of $82$ min. in the MRI laboratory using the protocols
reported in
\cite{A-H-M-P-P-S-V:RSSAMRI,A-A-H-J-K-K-L-M-M-P-S-V:LSDASMRIS}. We
obtained $107$ MRI scans which is only possible using well-optimised
experimental arrangements.  Of the $107$ scans, no more than $36$ were
prolonged vowels at $f_0\approx 104 \, \mathrm{Hz}$ (with sample
lengths $\approx 11.2 \, \mathrm{s}$) deemed usable for this study.
To obtain comparison data, same kind of speech recordings were carried
out in the anechoic chamber but neither the MRI coil reflections nor the
ambient noise were replicated.  Compared to MRI experiments, there are
no similar restrictions in the anechoic chamber, apart from test subject
fatigue. Thus, each vowel was now produced $10$ times since the larger
sample number was possible as a benefit of less demanding experimental
arrangement.

\section{MRI noise cancellation}
\label{CancellationSec}

We treat the measurement signals from speech and acoustic MRI noise
$s[t]$ and $n[t]$ for $t \in \{h, 2h, 3h, \ldots \}$ in their
digitised form where $h = 1/f_s$, and the sampling frequency $f_s
= 44\, 100 \, \textrm{Hz}$. The post-processing algorithm for these
discrete time signals is outlined in Fig.~\ref{AlgorithmNoiseFig}
(upper panel), and it consists of the following
Steps~\ref{LSQ}--\ref{SpecSub} that have been realised as MATLAB code:
\begin{enumerate}
\item \label{LSQ} {\bf LSQ:} Speech channel crosstalk is optimally
  removed from noise signal using coefficient $k$ from least squares
  minimisation.
\item \label{FreqResComp} {\bf Frequency response compensation:} The
  frequency response of the whole measurement system, shown in
  Fig.~\ref{AlgorithmNoiseFig} (upper panel), is compensated.  The
  peaks in the frequency response are due to the longitudinal
  resonances of the waveguides, used to convey the sound from inside
  the MRI scanner to the microphone array placed in a sound-proof
  Faraday cage.
\item {\bf Noise peak detection:} The noise power spectrum is computed
  by FFT, and the most prominent spectral peaks of noise are detected.
\item {\bf Harmonic structure completion:} The set of noise peaks is
  completed by its expected harmonic structure to ensure that most of
  the noise peaks have been found as shown in
  Fig.~\ref{AlgorithmNoiseFig} (lower panel on the left). There are
  heuristics involved so that the harmonics of the reference value of
  $f_0$ do not get accidentally removed. Details are described below
  in pseudocode.
\item \label{Chebyshev1} {\bf Notch filtering:} The noise peaks are
  removed by using notch filters provided by the MATLAB function
  \texttt{iircomb} with parameters \texttt{n} equal to the number of
  different harmonic overtone structures detected, and the $-3 \,
  \textrm{dB}$ bandwidth \texttt{bw} set at $6\cdot 10^{-3}$.
\item \label{SpecSub} {\bf Spectral subtraction:} A sample of the
  acoustic background (including, e.g., noise from the helium pump) of
  the MRI laboratory (without patient speech and scanner noise) is
  extracted from the beginning of the speech recording. Finally, the
  averaged spectrum of this ``silent sample'' is subtracted from the
  speech signal using FFT and inverse FFT; see \cite{SB:SANSUSS}.
\end{enumerate}

\begin{algorithm}
We associate with each spectral peak $p$ its location in spectrum $loc(p)$ in $Hz$, and its height $mag(p)$ in $dB$.
\caption{Adaptation to spectral structure}\label{euclid}
\begin{algorithmic}[1]
\State $P\gets$ set of all peaks found in the spectrum. 
\Procedure{FindHarmonics}{P}
\While {$P\neq \emptyset$}
\State $p \gets \max_{mag}P$
\State $P \gets P \setminus p$
\For {$q \gets P $ sorted by $|loc(p)-loc(P)|$}
\State $d \gets |loc(p)-loc(q)| $
\If {$d<cf_0$}
\State \textbf{continue}
\EndIf
\If {$\exists$ harmonics with fundamental $d$}
\State $F\gets F \cup $ \texttt{iircomb}$(f_s/d)$
\State $P \gets P \setminus \lbrace r \in P \,:\, r = n d, n \in \mathbb{Z} \rbrace$
\EndIf
\EndFor
\EndWhile
\State\Return {F}
\EndProcedure
\end{algorithmic}
Harmonics are considered successfully found at step 10, if $P$
contains four consecutive peaks with distance $d$.  The value $1.5$
has been used for the parameter $c$.
\end{algorithm}

The proposed approach differs essentially from the earlier approach
proposed in
\cite[Section~4]{A-A-H-J-K-K-L-M-M-P-S-V:LSDASMRIS}. Firstly, now
there is no direct time-domain subtraction of the measured noise
component from speech which makes the present approach more similar to
\cite{bresch}. For that reason, the low frequency components of speech
are not attenuated as a result of the proximity of recording sound
effect in dipole configurations.  Secondly, using notch filters
instead of high-order Chebyshev produces sharper removal of unwanted
spectral components \RevisionText{ with much reduced musical noise
  artefact compared to what was reported in
  \cite{A-A-H-J-K-K-L-M-M-P-S-V:LSDASMRIS}. The comb filter is a more
  efficient way of removing higher harmonics of spectral peaks in the
  entire spectrum. In the current approach, the filter degree is
  determined by the Nyquist frequency $f_s/2 = 22\, 050\, \mathrm{Hz}$
  and the number of notches required, making the computations much
  less intensive. However, using Chebyshev filters made it possible to
  vary the bandwidth of the stop bands as a function of frequency
  which possibility is now lost.}

\RevisionText{In \cite{A-A-H-J-K-K-L-M-M-P-S-V:LSDASMRIS}, the
  post-processed speech recordings during MRI were classified with
  linear discriminant classifier, using the speech recorded in the
  anechoic chamber as a learning set. This experiment yielded 62\%
  correct classifications.  Repeating the experiment using the same
  speech data, the improved post-processing algorithm, and better
  accounting for the strong exterior resonance at $\approx 1kHz$ as
  discussed in Section~\ref{IdentificationSec} below, the proportion
  of correctly classified vowels increases to 72\%. Further
  significant improvement in classification accuracy does not seem
  possible since a strong systematic component is present in
  classification errors of both classification experiments, reflecting
  the properties of the speech data. More precisely, many
  \textipa{[\ae]} get classified as \textipa{[e]}, and many
  \textipa{[e]} get classified as \textipa{[i]}. Looking at the
  spectral envelopes of \textipa{[\ae]} in
  Fig.~\ref{SpectralEnvelopesFront}, two different kinds of behaviour
  can be seen in the upper curves.  Based on only $F_1$ and $F_2$,
  samples with the lower first peak location (i.e.,
  $F_1[$\textipa{\ae}$]$) are almost indistinguishable from
  \textipa{[e]} recorded in the anechoic chamber. This results in the
  first kind of systematic error. The second type of error is due to
  the systematic overestimation of $F_2[\textipa{e}] \approx 2 \,
  \mathrm{kHz}$ in speech recorded during MRI as can be seen in
  Fig.~\ref{VowelDiscrepancy}. This artefact is connected to the
  acoustics inside the MRI head coil in
  Sections~\ref{FormSubSec}~and~\ref{IdentificationSec}.}

\section{Performance analysis}
\label{PerfSec}

\subsection{Validation through synthetic signals}
\label{ValSubSec}

\begin{figure}
\begin{center}
  \includegraphics[width=0.23\textwidth]{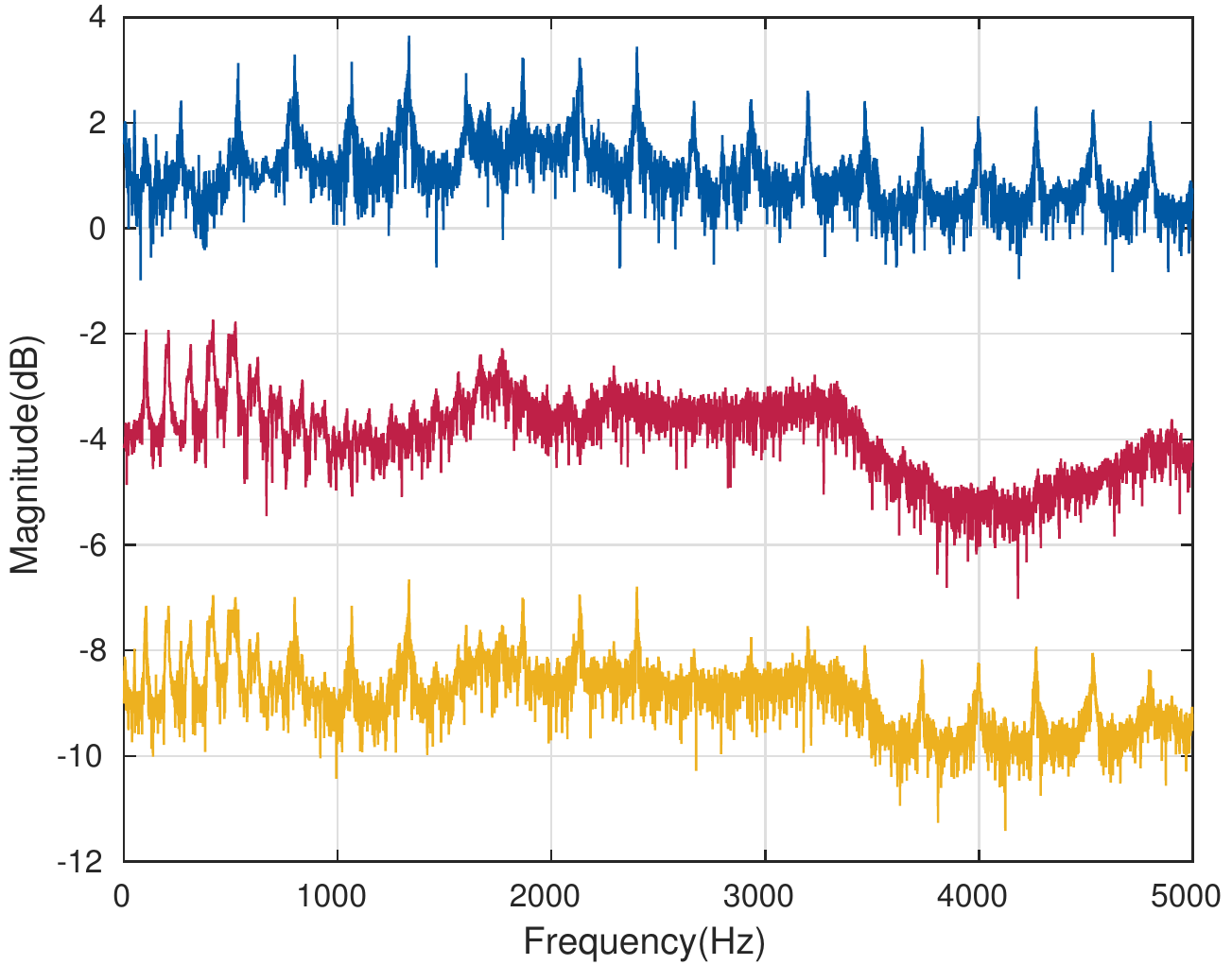}
  \includegraphics[width=0.23\textwidth]{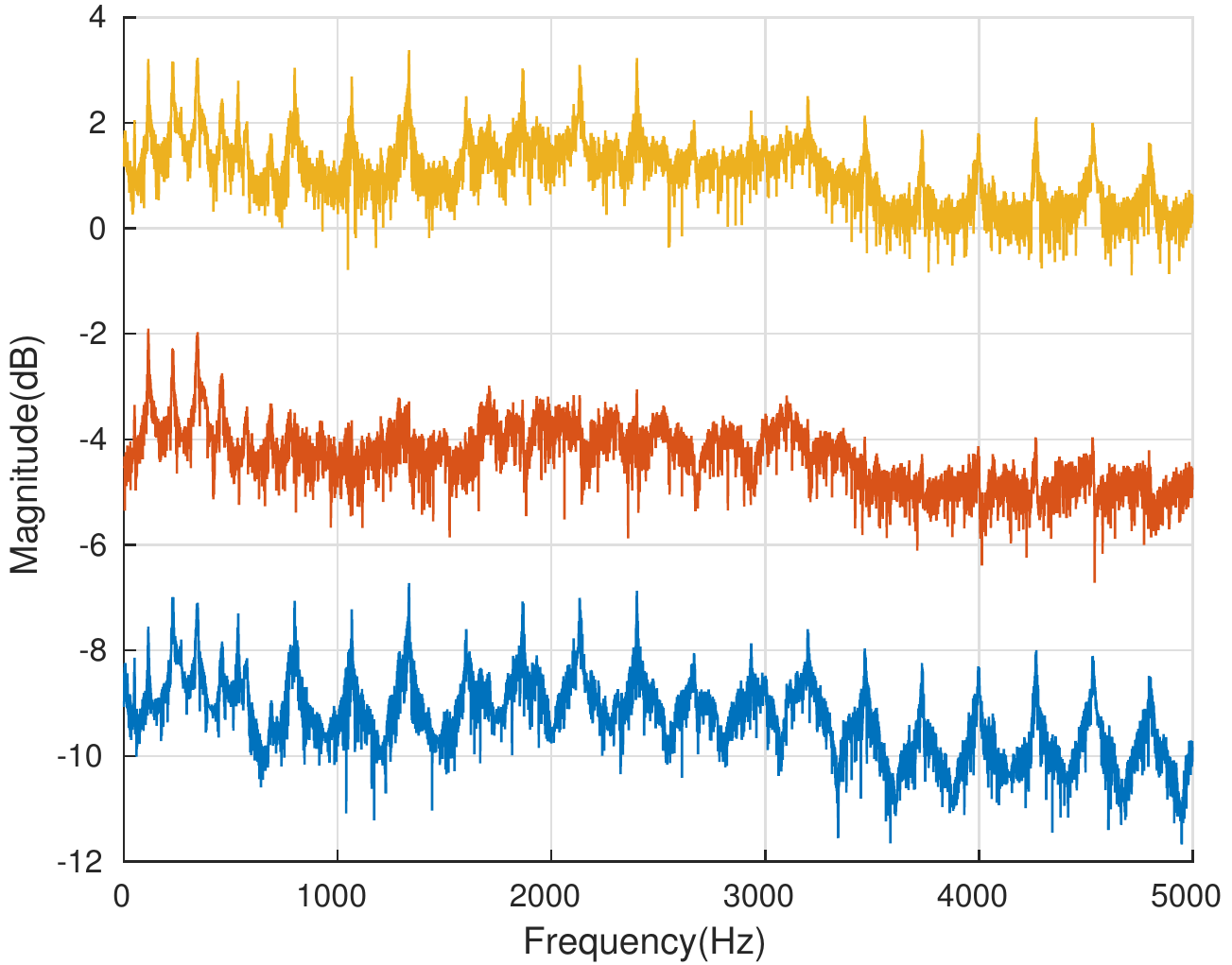}
\end{center}
\caption{\label{ValidationSignals} Illustration of the artificially
  noise-contaminated vowel signal.  On the left, MRI noise (upmost),
  pure vowel signal (middle), and the synthetic signal as their sum
  (lowest).  On the right, synthetic signal (upmost), signal after
  post-processing using the proposed algorithm (middle), and the
  reconstructed noise (lowest). 
}
\end{figure}

The formant extraction from noisy speech can validated using
artificially noise contaminated speech where the original formant
positions are known precisely. Pure vowel signals were taken from
comparison data for each vowel in \textipa{[\textscripta , e, i, o, u,
    y, \ae , \oe]}, and their formants $F_1, F_2$, and $F_3$ were
computed\footnote{Throughout this article, the MATLAB function {\tt \,
    arburg} is used for producing low-order rational spectral
  envelopes from which the formants are extracted by locating poles.}.
A sample of MRI noise (without any speech content) was recorded using
the experimental arrangement detailed in
\cite[Section~3]{A-A-H-J-K-K-L-M-M-P-S-V:LSDASMRIS}, and it was mixed
with each vowel sample so that the speech and noise components have
equal energy contents ($\mathrm{SNR} \approx 0 \, \mathrm{dB}$).  The
post-processing algorithm described in Section~\ref{CancellationSec}
was then applied to these signals, of which an example is shown in
Fig.~\ref{ValidationSignals}.  

It was first  observed that the post-processing increases the SNR of the
artificially noise-contaminated signals by $9 \ldots 14 \,
\mathrm{dB}$ depending on the vowel.  The three formants $F_1, F_2$,
and $F_3$ were extracted from artificially noise contaminated vowels
after they had been post-processed. The resulting formant frequencies
are within $-0.5 \ldots 0.3$ semitones from those measured from the
original pure vowels, except for the outlier $F_2$[\textipa{o}] where
the discrepancy is $1.1$ semitones.

\begin{table}[h]
 \centerline{
    \begin{tabular}{|c|c|c|c|}
\textbf{Vowel}& $F_1$ & $F_2$ & $F_3$ \\
\hline
{[\textipa{\textscripta}]} &598 &1094   & 1918  \\
{[\textipa{e}]} &453  & 1691 & 2255  \\
{[\textipa{i}]} & 318   & 1900& 2097  \\
{[\textipa{o}]} &465  &815& 2233  \\
{[\textipa{u}]} & 410 &898 & 1934 \\
{[\textipa{y}]} & 379  &1535 & 2034 \\
{[\textipa{\ae}]} &562  & 1452 & 2375  \\
{[\textipa{\oe}]} &436  &1400  &2076 \\
\end{tabular}
    \begin{tabular}{|c|c|c|c|}
\textbf{Vowel}& $F_1$ & $F_2$ & $F_3$ \\
\hline
{[\textipa{\textscripta}]} &615 &1129   & 2021  \\
{[\textipa{e}]} &443  & 1714 & 2299  \\
{[\textipa{i}]} & 327   & 1909 & 2293  \\
{[\textipa{o}]} &451  &858& 2088  \\
{[\textipa{u}]} & 416 &921 & 2041 \\
{[\textipa{y}]} & 390  &1533 & 2015 \\
{[\textipa{\ae}]} &559  & 1476 & 2319  \\
{[\textipa{\oe}]} &428  &1421  &2099 \\
\end{tabular}
}

\caption{Original formants (left) and formants extracted after the
  artificial addition of MRI noise and subsequent noise cancellation
  (right).}
\end{table}

The average formant discrepancies of under $2.8$ semitones were
reported in \cite[Table~3]{A-A-H-J-K-K-L-M-M-P-S-V:LSDASMRIS} between
speech formants and Helmholtz resonances computed from vocal tract
geometries (without any model for the surrounding space) that were
obtained by simultaneous MRI.  Also, the observations in
\cite{A-M-V-S-P:EMAEMRID} provide magnitudes for formant error that
results from inherent variation in long vowel productions due to test
subject adaptation and fatigue. Comparing these values with the
results on artificially contaminated speech, we conclude that formant
extraction from algorithmically post-processed signals can be regarded
as a relatively small error source. 

\subsection{Comparison of spectral tilts}
\label{TiltSec}
 In addition to formants, another important spectral characteristic of
 speech signals is the \emph{spectral tilt} or \emph{roll-off}.  It is
 a measure of attenuation at higher frequencies that are still
 relevant to speech. We quantify the spectral tilt by first fitting a
 low-order rational spectral envelope on the frequency range of
 speech, and then finding the LSQ regression line to the envelope on
 the logarithmic frequency range between $465 \, \mathrm{Hz}$ and $5
 \, \mathrm{kHz}$. The bound $465 \, \mathrm{Hz}$ is the mean of all
 $F_1$'s present in the dataset.

\begin{table}[h]
 \centerline{
    \begin{tabular}{|l|c|c|c|c|c|c|c|c|}
\hline
&{[\textipa{\textscripta}]}  &{[\textipa{e}]}  &{[\textipa{i}]}  &{[\textipa{o}]} &{[\textipa{u}]} &{[\textipa{y}]} &{[\textipa{\ae}]} &{[\textipa{\oe}]} \\
\hline
\textbf{Anech} &12.2 &11.9 &9.0 &14.5 &15.6 &12.6 &11.3 &12.7  \\
\textbf{MRI} &15.7 &13.9 &9.2 &17.9 &15.3 &13.5 &14.0  & 15.2 \\
\hline 
\end{tabular}}
\caption{\label{TiltTable} Spectral tilts (in $\mathrm{dB}/\mathrm{octave}$) from
  recordings in the anechoic chamber and from samples recorded during
 the  MRI noise after post-processing. }
\end{table}

\begin{figure}[t]
\begin{center}
  \includegraphics[height=4.8cm ]{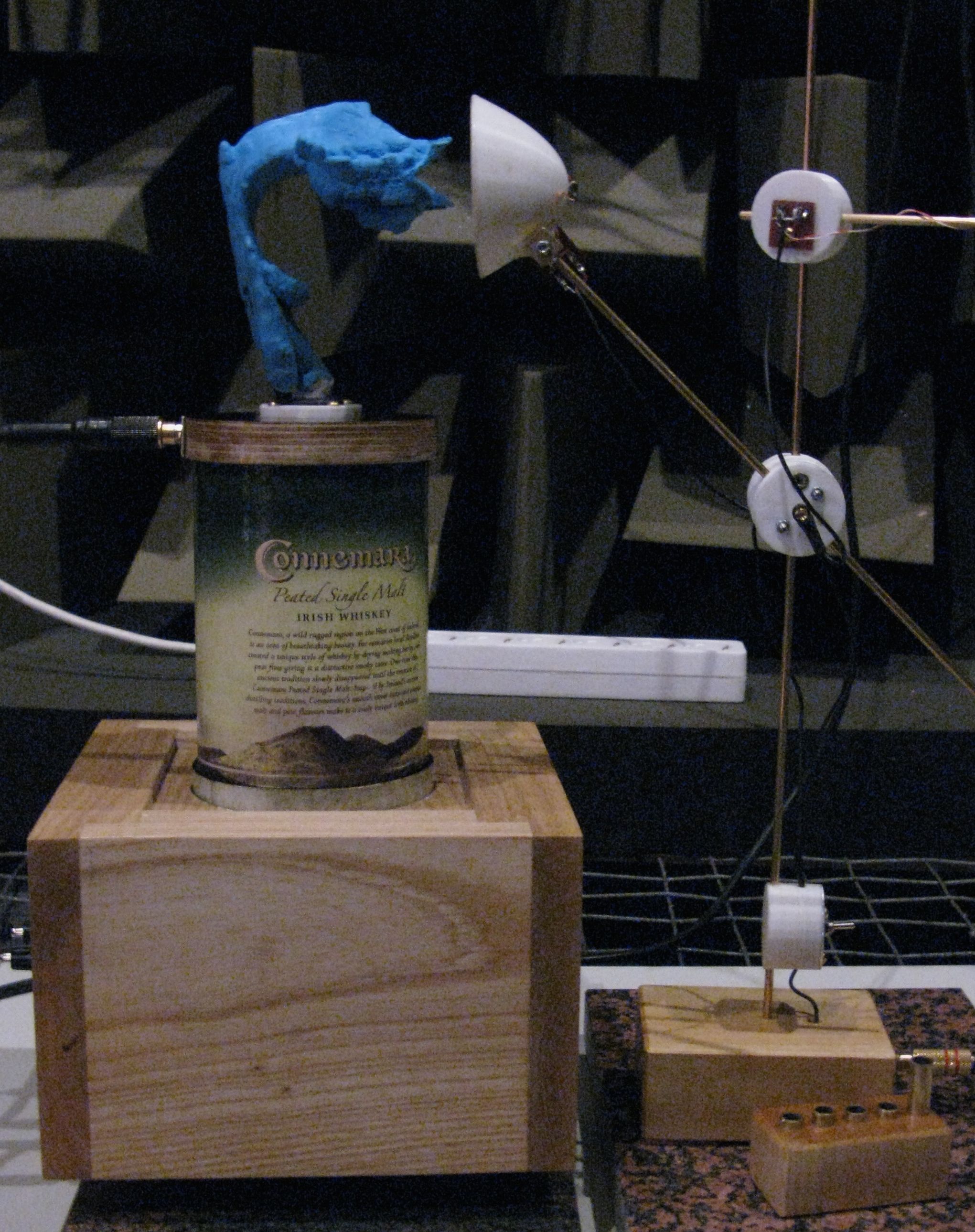}
\vspace{0.2cm}
  \includegraphics[height=3.6cm, angle =90]{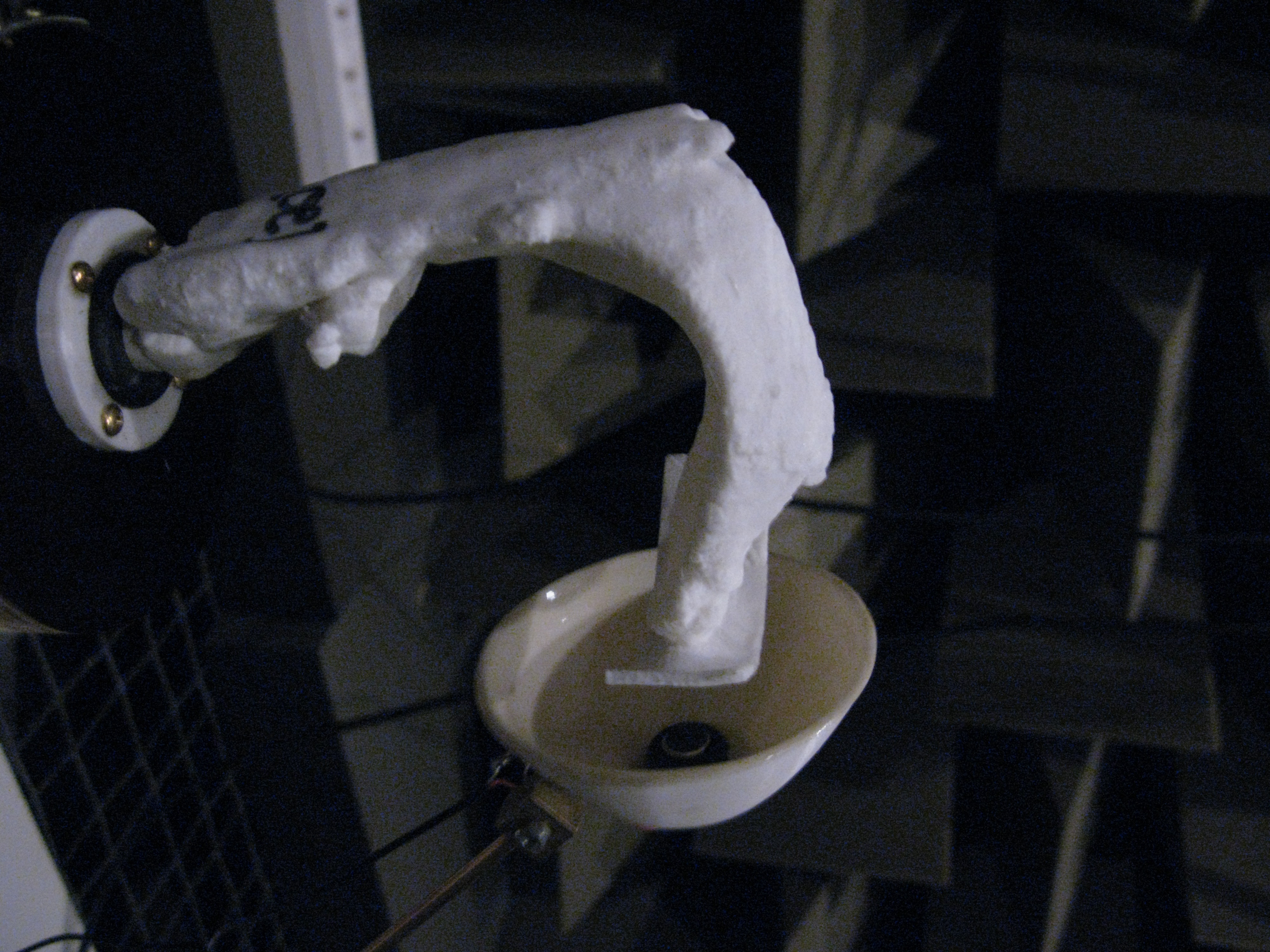}
\end{center}
  \caption{\label{InstrumentDetail} A detail of the sweep measurement
    arrangement for 3D printed vocal tract configurations of
    [\textipa{\textscripta}, \oe].}
\end{figure}

The spectral tilt data is given in Table~\ref{TiltTable}.  The
roll-off in post-processed speech during the MRI is systematically larger
than in comparison data (in average by $1.9 \, \mathrm{dB}$), the only
exception being the vowel [\textipa{y}]. We point out that the two
kinds of spectral tilt data in Table~\ref{TiltTable} correlate
strongly ($R = 0.78$). As can be seen from Fig.~\ref{Sweeps} (last
panel), the difference of the average spectral tilts is quite
small. The difference is partly explained by the fact that there was a
lot of more attenuating material around the test subject in the MRI
scanner, compared to experiments in the anechoic chamber.

\subsection{Comparison to sweeps in physical models}
\label{3DPrintSec}

Three of the MR images corresponding to Finnish quantal vowels
[\textipa{\textscripta, i, u}] were processed into 3D surface models
(i.e., STL files) and intersectional area functions for Webster's
equation as explained in \cite{A-H-H-K-M-S-R:ASEMRIDMHVT}. Fast
prototyping was used to produce physical models from the STL files in
ABS plastic with wall thickness $2 \, \mathrm{mm}$.  The printed
models extend from the glottal position to the lips, and they were
coupled to a custom acoustic source (see Fig.~\ref{InstrumentDetail})
whose design resembles the loudspeaker-horn construction shown in
\cite[Fig.~1]{T-K-E-S-W:EEIFPSKGFTC}; see also
\cite{T-M-K:AAVTDVPFDTDM}.

\begin{figure}[t]
\begin{center}
  \includegraphics[width=0.22\textwidth]{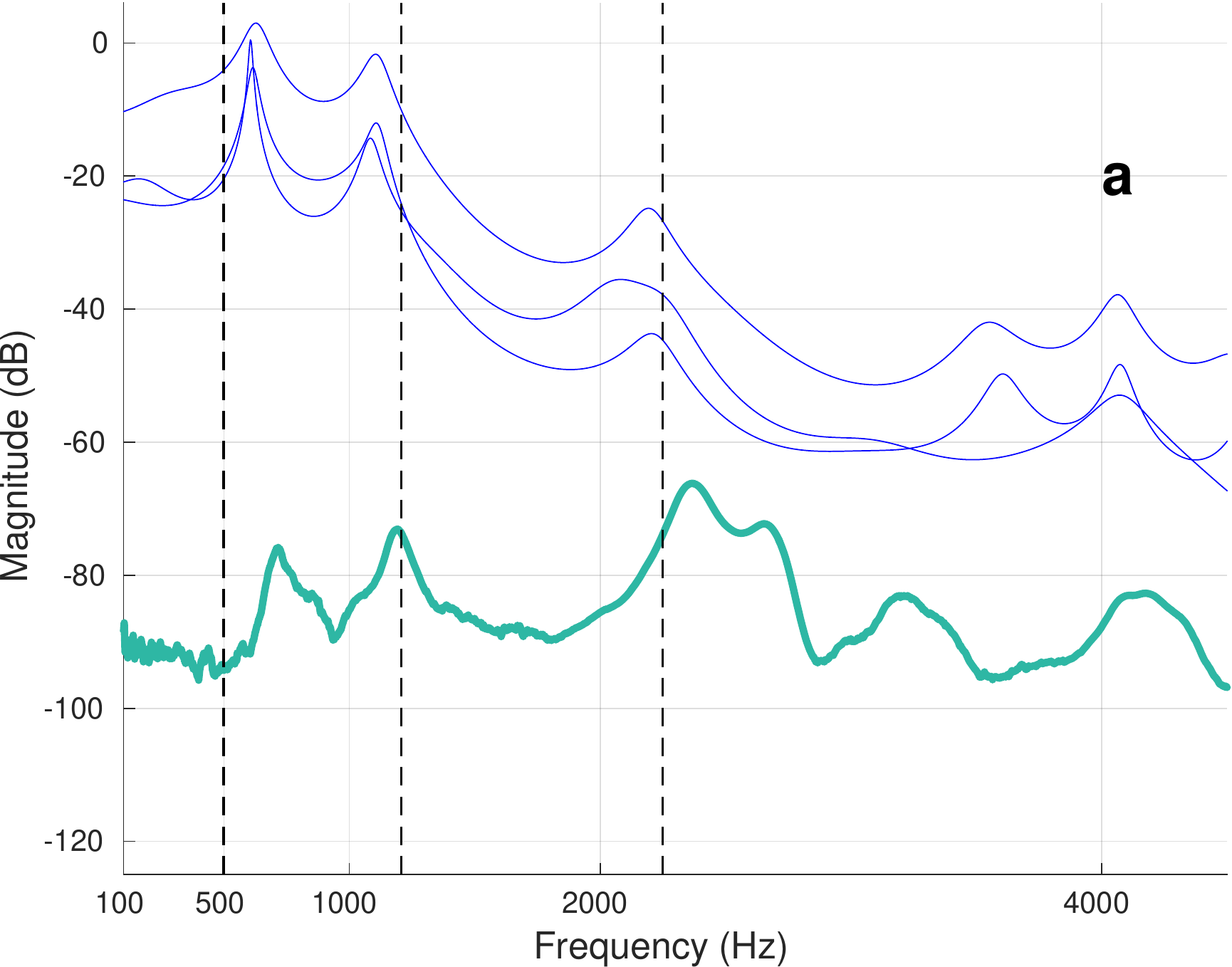}
  \includegraphics[width=0.22\textwidth]{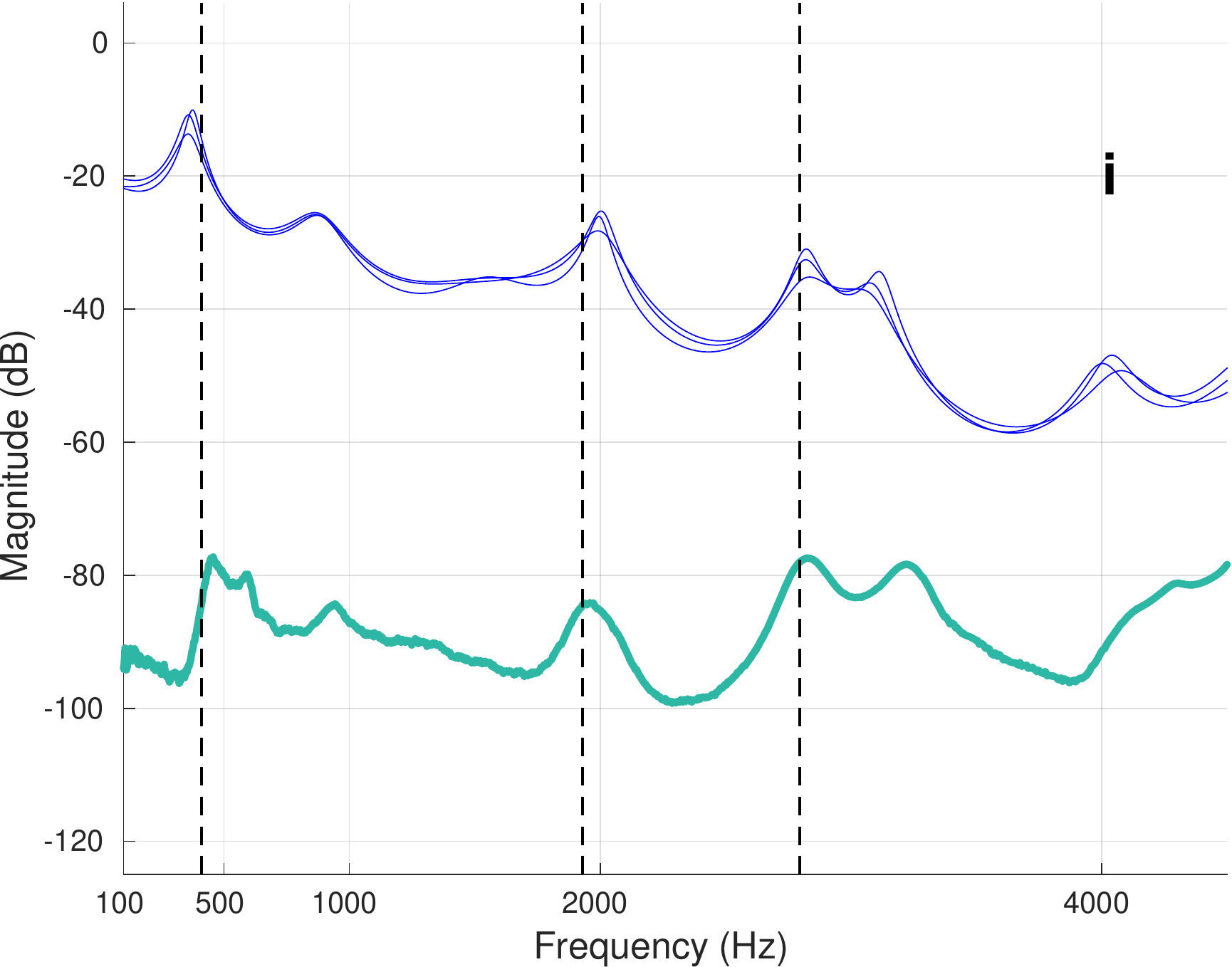}
\end{center}
\begin{center}
  \includegraphics[width=0.22\textwidth]{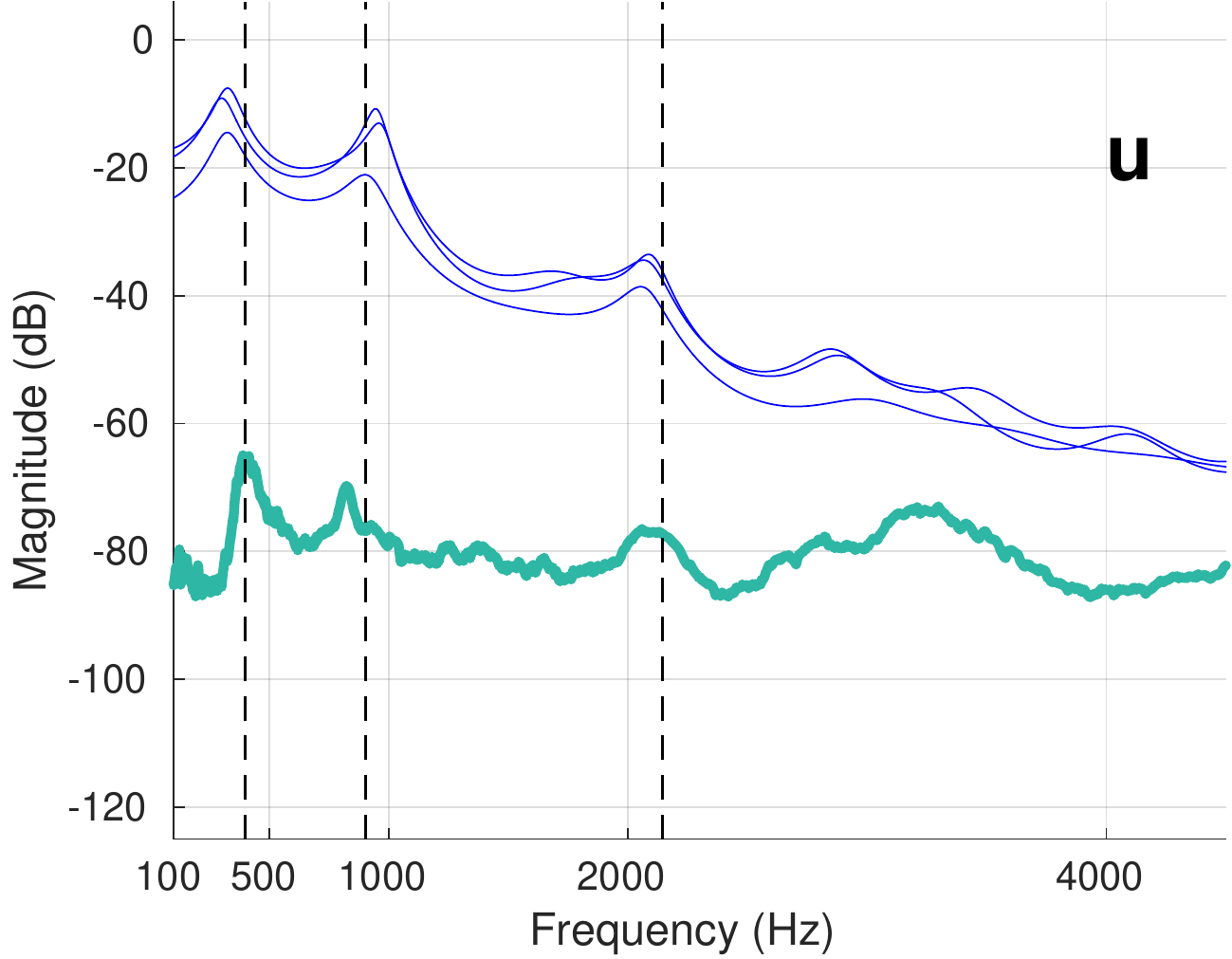}
  \includegraphics[width=0.22\textwidth]{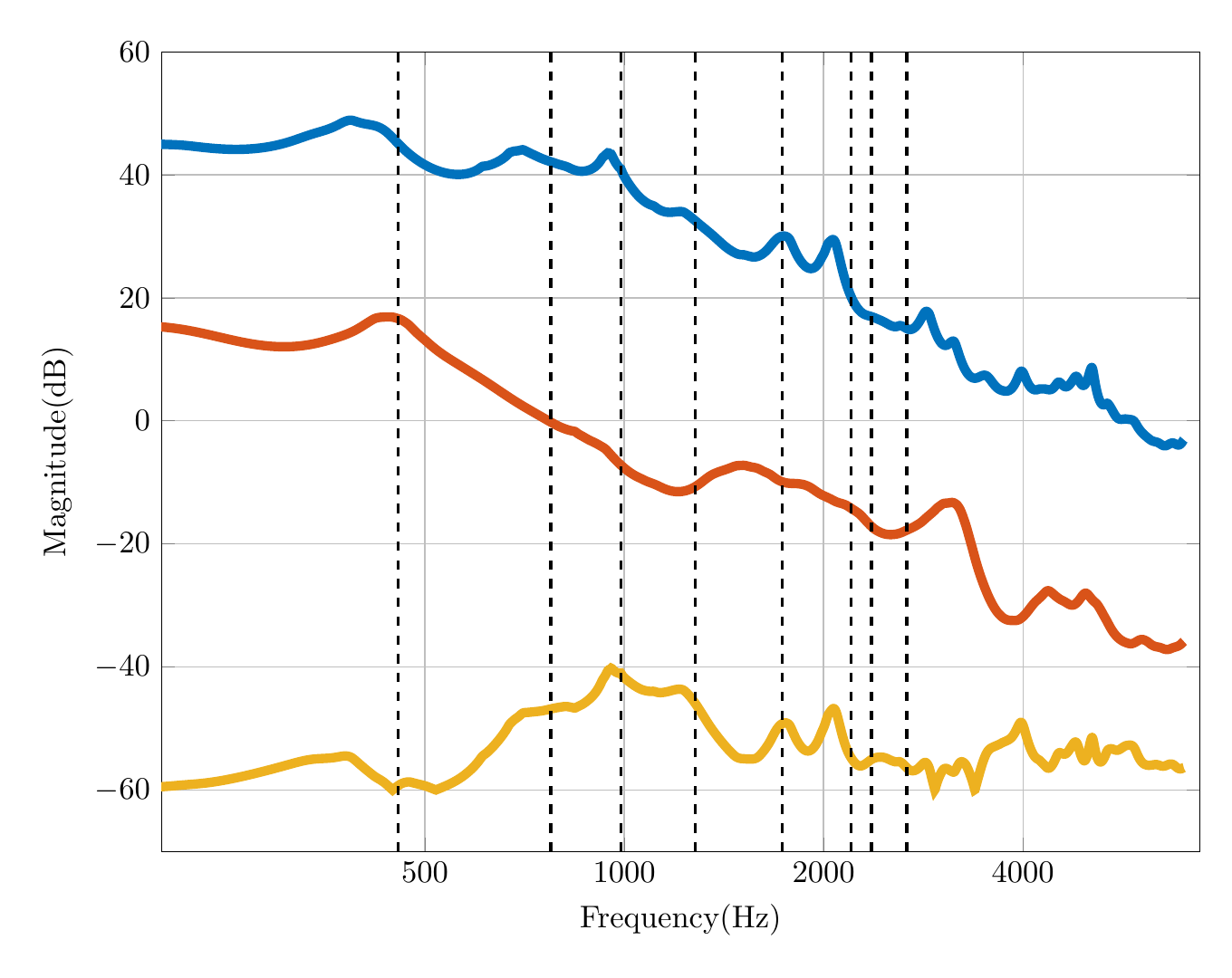}
\end{center}
  \caption{\label{Sweeps} The first three panels: Spectral envelopes
    and computationally obtained resonances of [\textipa{\textscripta,
        i, u}]. The upper curves are power spectral densities of
    speech recorded during an MRI scan.  The lower curves are
    frequency responses measured from the physical models that have
    been produced from the MR images.  The vertical lines indicate the
    three lowest resonances computed by Webster's model from the same
    VT geometry using the mouth impedance optimisation process
    introduced in \cite{K-K-M-O:MIOVTRV}. The last panel: Averages of
    spectral envelopes of Finnish vowels \textipa{[\textscripta , e,
        i, o, u, y, \ae , \oe]} from two different kind of
    recordings. Each vowel appears in the averages with the same
    weight. The topmost curve describes speech recorded during the MRI
    scan, the middle curve recordings in the anechoic chamber, and the
    lowest curve is their difference. The averaging highlights the
    common features (partly due to the exterior acoustics) within both
    kinds of vowel recordings.  The vertical dashed lines represent
    $k$-means cluster centroids of the Helmholtz resonant frequencies
    computed using a 3D model of the MRI head coil.}
\end{figure}

The acoustic source contains an electret (reference) microphone
($\oslash \, 9 \, \mathrm{mm}$, biased at $5 \, \mathrm{V}$) at the
glottal position, and another similar (signal) microphone was placed
near the lips. 
A sinusoidal logarithmic sweep was preweighted by the iteratively
measured inverse response of the acoustic source in order to obtain a
uniform sound pressure level at the reference microphone for all
frequencies of interest. The frequency responses of the physical
models (and reference resonators with known resonant frequencies were
measured using this arrangement between $80 \, \mathrm{Hz} \ldots 7 \,
\textrm{kHz}$.

As can be seen from Fig.~\ref{Sweeps}, there is good correspondence
between the spectra of de-noised speech from MRI experiments and the
spectra from physical models of the simultaneously imaged vocal tract
geometry. There are some extra peaks in both kinds of spectra that
correspond to spurious resonances not due to the vocal tract geometry.
We point out that the physical models did not contain the face, and
the sweep measurements were carried out in an open acoustic
environment in the anechoic chamber. This is in contract to the speech
recordings that were carried out within MRI head and neck coils
\cite{A-H-M-P-P-S-V:RSSAMRI,A-A-H-J-K-K-L-M-M-P-S-V:LSDASMRIS}.

It is worth observing from Fig.~\ref{Sweeps} that the spectral tilt (as
defined in Section~\ref{TiltSec}) of the frequency response from
physical models is practically $0 \,\mathrm{dB}/\mathrm{octave}$. This
is due to two reasons: \textrm(i) A 3D printed vocal tract is a
virtually lossless acoustic system apart from the radiation losses
through mouth opening, and \textrm(ii) the glottal excitation in
natural speech has its characteristic roll-off of $11 \ldots 16
\,\mathrm{dB}/\mathrm{octave}$ whereas the measurements from the
physical models were carried out keeping the sinusoidal sound pressure
constant at the glottal position.

\subsection{Perceptual evaluation}
\label{PerSec}

A listening experiment was carried out to evaluate the effect of
post-processing on vowel recognition. In the experiment, $12$ subjects
(of which two were female) listened to $48$ recordings of vowel
phonation. The recordings consisted of $6$ samples of each Finnish
vowel in [\textipa{\textscripta, e, i, o, u, \ae, \oe}]; half of the
samples were unprocessed recordings from the anechoic chamber ($24$ in
total, three for each vowel), while the rest had undergone the MRI
noise contamination and de-noising process described in
Section~\ref{ValSubSec}.  The duration of each sample was $10 \,
\mathrm{s}$.

\begin{table}[h]
\begin{center}
\textbf{a) Vowel samples from anechoic chamber}\\
\begin{tabular}{lcccccccc}
\toprule
& \multicolumn{8}{c}{\textbf{categorised as}} \\
\cmidrule(lr){2-9}
\textbf{target} & [\textipa{\textscripta}] & [\textipa{e}] & [\textipa{i}] & [\textipa{o}] & [\textipa{u}] & [\textipa{y}] & [\textipa{\ae}] & [\textipa{\oe}]\\
\midrule
$[$ \textipa{\textscripta} $]$ & 36& 0 &0 & 0 &0 &0 &0  &0  \\
$[$ \textipa{e} $]$ &0 &33  &0 & 0 &0 &0 &0  &3  \\
$[$ \textipa{i} $]$ & 0& 0 &36 &0  &0 &0 &0  &0  \\
$[$ \textipa{o} $]$ & 6& 0 &0  & 30  &0 & 0& 0 &0  \\
$[$ \textipa{u} $]$ & 0 & 0 & 0 &13  &23& 0& 0 & 0 \\
$[$ \textipa{y} $]$  & 0& 0 & 0& 0 &0 & 32&  0& 4 \\
$[$ \textipa{\ae} $]$ & 0& 1 &0 &  0&0 &0 &32  & 1 \\
$[$ \textipa{oe} $]$ &0 &3  &0 &0  & 0&0 & 0 & 33 \\
\bottomrule
\end{tabular}

\vspace{7mm}

\textbf{b) Artificially MRI noise contaminated samples}\\
\begin{tabular}{lcccccccc}
\toprule
& \multicolumn{8}{c}{\textbf{categorised as}} \\
\cmidrule(lr){2-9}
\textbf{target} & [\textipa{\textscripta}] & [\textipa{e}] & [\textipa{i}] & [\textipa{o}] & [\textipa{u}] & [\textipa{y}] & [\textipa{\ae}] & [\textipa{\oe}]\\
\midrule
$[$ \textipa{\textscripta} $]$ &36 &0  &0 & 0 & 0& 0&0  &0  \\
$[$ \textipa{e} $]$ & 0&30  &0 &0  &0 &0 & 0 & 6 \\
$[$ \textipa{i} $]$ &0 & 0 &36 &0  &0 &0 & 0 &0  \\
$[$ \textipa{o} $]$ & 8& 0 &0 &28  &0 &0 &0  &0  \\
$[$ \textipa{u} $]$ &0 & 0 & 0& 15 &21 &0 & 0 &0  \\
$[$ \textipa{y} $]$ & 0 &0  &0 &0  &0 & 27& 0 & 9 \\
$[$ \textipa{\ae} $]$ &0 & 0 &0 & 0 &0 &0 & 36 &0  \\
$[$ \textipa{oe} $]$ &0 & 0 &0 &1  &0 &0 &0  &35  \\
\bottomrule
\end{tabular}

\end{center}
\caption{\label{PerceptualTable} Results of the perceptual comparison
  experiment on vowels, some of which were artificially contaminated
  by MRI noise and then de-noised. Quite many target samples of
  [\textipa{u}] were classified as [\textipa{o}] in both kinds of
  samples.}
\end{table}

The test subjects were allowed to listen each sample as many times as
they wanted. Using a computer interface, they reported the vowel that
the phonation resembled the most in their opinion. The results of the
perceptual experiment are given in Table~\ref{PerceptualTable}. As a
conclusion, there is a slight increase in classification mistakes
induced by the proposed algorithm, but the increase is a fraction of
the classification mistakes due to natural speech variation in the
samples used.  To draw statistically significant conclusions on such
small effects would require a considerably larger data set.

\section{Formant extraction from noisy speech}
\label{FormSubSec}

After four validation experiments on the post-processing algorithm
described in Section~\ref{CancellationSec}, it is time to apply it on
true speech data, recorded during an MRI scan. Our purpose is to show
by comparative studies that the acoustic environment in the MRI
scanner introduces resonant artefacts to speech signals that are large
enough to be clearly quantifiable using the proposed algorithm.

To increase the number of vowel sound samples from MRI experiments,
six partial samples of $1 \, \mathrm{s}$ were taken from each
recording.  These partial samples are separated from each other by at
least $1 \, \textrm{s}$ of time to enhance the independence of the
samples. This sixfold increase of the original sample number improves
the statistical analysis given in Table~\ref{PValues}. Spectral
envelopes of all speech samples are shown in
Fig.~\ref{SpectralEnvelopesFront} where variance between same vowel
productions in different MRI scans (or different parts of the same
scan) can be observed.

We proceed to show that some of the extracted formant means of samples
from the anechoic chamber and the MRI laboratory are significantly
nonequal. The estimated formant means $\mu_{ac}$ and $\mu_{mri}$ are
compared using Student's t-distribution where the degrees-of-freedom
is determined by the Smith-Satterwaithe procedure; see the unequal
variance test statistics in, e.g., \cite[Section~10.4]{M-A:IPSAS}.  In
case of the vowel formant $F_j$\textipa[\textscripta] for $j = 1, 2,
3$, our null hypothesis is that
\begin{equation*}
  H_0: \mu_{ac}\left (F_j \text{[\textipa{\textscripta}]} \right ) = \mu_{mri} \left ( F_j [\text{\textipa{\textscripta}]} \right )
\end{equation*}
\begin{figure}
\begin{center}
  \includegraphics[width=0.48\textwidth]{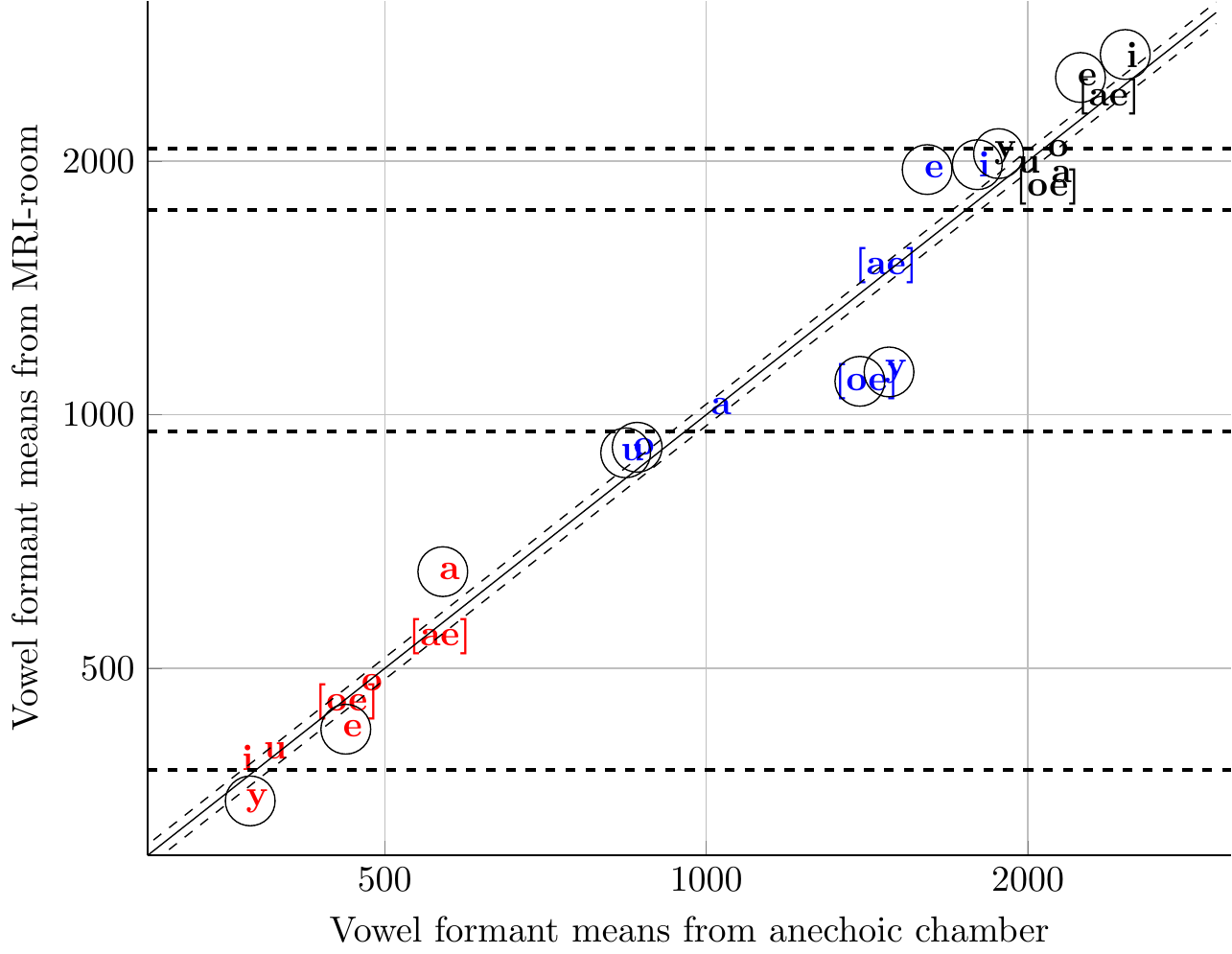}
\end{center}
\caption{\label{VowelDiscrepancy} Estimates of formants $F_1$, $F_2$,
  and $F_3$ that have been extracted from the vowel samples of
  \textipa{[\textscripta , e, i, o, u, y, \ae , \oe]} recorded during
  the MRI.  They are plotted against the comparable data recorded in
  the anechoic chamber from the same test subject. The diagonal dashed
  lines describe the error bounds of $\pm 0.5$ semitones as obtained
  in Section~\ref{ValSubSec}.  Where the formant discrepancy is
  statistically significant at $p \geq 0.95$, the vowel has been
  encircled; see Table~\ref{PValues}. The horizontal dashed lines show
  peaks of the spectral envelopes in Fig~\ref{Sweeps} (last panel)
  that were identified as resonances external to the vocal tract.}
\end{figure}

We try to reject $H_0$ by showing that its converse $H_1$ is true with
high probability, say $p > 0.95$, in which case the experiment
indicates that the formant extraction from the two data sources is not
consistent. The results of the experiments are given in
Table~\ref{PValues} where the $p$-values are given.  We conclude that
$H_0$ gets typically rejected for $F_2$ in all vowels except
[\textipa{\textscripta, o, \ae}] and for all formants in vowels
[\textipa{e, i}].

The formant means from post-processed speech during the MRI are plotted in
Fig.~\ref{VowelDiscrepancy} against their counterparts recorded in
the anechoic chamber from the same test subject. If these two datasets
were perfectly consistent, all data points would be expected to appear
between the two diagonal dashed lines, representing the maximum error
of formant extraction from noisy speech as discussed in
Section~\ref{ValSubSec}. We conclude that (at least) 12 of the
discrepancies shown in Fig.~\ref{VowelDiscrepancy} reflect actual
differences of the speech data recorded in MRI laboratory, compared to
similar data from the anechoic chamber.

It is worth observing that the formant discrepancy in
Fig.~\ref{VowelDiscrepancy} shows a peculiar staircase pattern where
two plateaus appear near $1 \, \textrm{kHz}$ and $2 \, \textrm{kHz}$.
More precisely, we observe that in samples recorded during the MRI, we
have $F_2$[\textipa{y}], $F_2$[\textipa{\oe}] $\to 1 \, \mathrm{kHz}$
from above and $F_2$[\textipa{e}], $F_2$[\textipa{i}] $\to 2 \,
\mathrm{kHz}$ from below.  The vertical level at $1 \, \textrm{kHz}$
coincides with an extra peak appearing in
Fig.~\ref{SpectralEnvelopesFront} in most of spectral envelopes of
signals recorded during the MRI; notable exceptions are the vowels
[\textipa{\textscripta},u,o] where $F_2 \approx 1 \, \textrm{kHz}$
would conceal any extra peak. These extra peaks can also be seen in
Fig.~\ref{Sweeps} (last panel) where the spectral envelopes of all
vowel recordings in the MRI laboratory (in the anechoic chamber,
respectively) have been averaged to downplay the vowel specific
formant peaks.
It has been excluded by frequency response measurements and ensuing
equalisation that these peaks could be an artefact of the speech
recording instrumentation.

\begin{table}[h]
  \centerline{
    \begin{tabular}{|c|c|c|c|c|c|c|c|c|}
      \hline  
        & [\textipa{\textscripta}] & [\textipa{e}] & [\textipa{i}] & [\textipa{o}] & [\textipa{u}] & [\textipa{y}] & [\textipa{\ae}] & [\textipa{\oe}]    \\
      \hline   
       $F_1$ &\textbf{0.99} & \textbf{0.98} & 0.84          &   0.14          &   0.70         &  \textbf{0.95} & 0.25  & 0.07  \\
       $F_2$ & 0.21         & \textbf{0.99} & \textbf{0.99} &   \textbf{0.99} & \textbf{0.98}  & \textbf{0.99}  & 0.81  & \textbf{0.98}\\
       $F_3$ & 0.82         & \textbf{0.99} & \textbf{0.99} &   0.60          & 0.17           &  \textbf{0.99}          & 0.61  &0.75 \\
      \hline   
  \end{tabular}}
  \caption{\label{PValues} The $p$-values computed with Smith-Satterwaith
    procedure for distributions with unequal variances. Formant
    samples that reject the null hypothesis $H_0$ at $p > 0.95$ are
    written in bold.}
\end{table}

 A similar staircase pattern to
Fig.~\ref{VowelDiscrepancy} near frequencies $1 \, \textrm{kHz}$ and
$2 \, \textrm{kHz}$ has been observed in \cite[Chapter~5,
  Fig.~5.4]{AK:AVTMRISM} where measured formant and computed resonance
pairs have been plotted against each other. The vocal tract resonances
in \cite{AK:AVTMRISM} have been computed by the Helmholtz equation from
MRI data without exterior space modelling, and the formants have
extracted from recordings during the MRI as explained in
\cite[Section~5]{A-A-H-J-K-K-L-M-M-P-S-V:LSDASMRIS}.

\section{Identification of exterior resonances}
\label{IdentificationSec}

The statistically significant discrepancy in
Fig.~\ref{VowelDiscrepancy} is expected to be a combination of three
different sources: \textrm{(i)} Perturbation\footnote{The discrepancy
  in vowel formants extracted from speech may be due to
  misidentification of exterior formants as adjacent vocal tract
  formants, or there may be ``frequency pulling'' of a correctly
  identified vowel formant by an adjacent exterior formant. In
  Helmholtz computations, we can always tell the true formants by
  looking at the corresponding pressure eigenmodes. Only spectrogram
  data is available from measured speech.} of the vocal tract
resonances by the adjacent exterior space resonances, caused by
reflections from test subject's face and MRI head coil surfaces;
\textrm{(ii)} Lombard speech due to the acoustic noise during the MRI
(see \cite{hazan2012clear,vainio2012lombard}); and \textrm{(iii)}
active adaptation of the test subject to the constrained space
acoustics inside the MRI head coil. \RevisionText{Of these three
  possible partial explanations, only the first can be studied without
  carrying out extensive experiments with test subjects.  Instead, we
  can use the simultaneously obtained MR image of the vocal tract for
  numerical resonance computations in order to investigate the
  acoustic artefacts in speech caused by the MRI coil.}

\begin{figure}[h]
\begin{center}
  \includegraphics[width=0.20\textwidth]{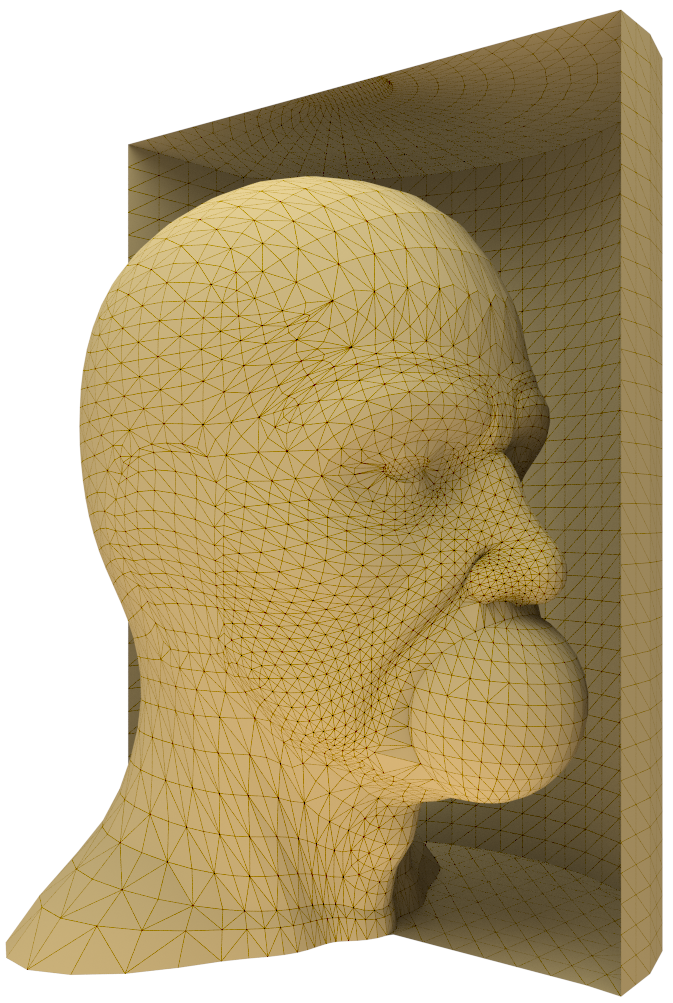}
  \includegraphics[width=0.23\textwidth]{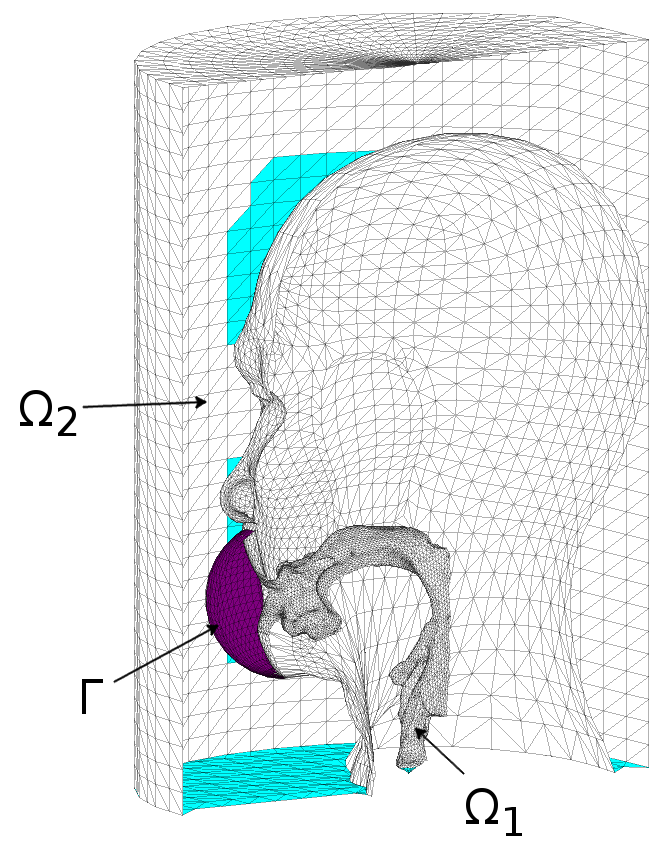}
\end{center}
\begin{center}
  \includegraphics[width=0.18\textwidth]{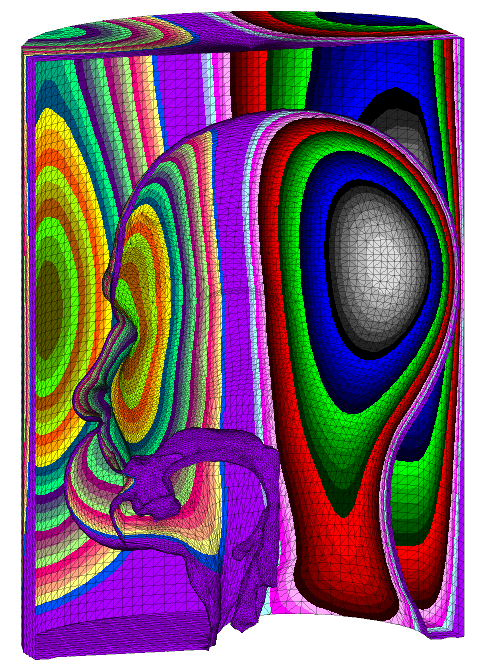}
\end{center}
\caption{\label{exterior} Top panels: An illustration of the
  computational domains used for identifying the acoustic resonances
  within MRI head coil.  The computational domains $\Omega_1$,
  $\Omega_2$, and the interface $\Gamma$ are shown on the right.
  Bottom panel: The modal pressure distribution at the domain boundary
  at the resonant frequency $1062 \, \mathrm{Hz}$.}
\end{figure}
\begin{figure}[t]
\begin{centering}
\includegraphics[width=40mm,height=30mm]{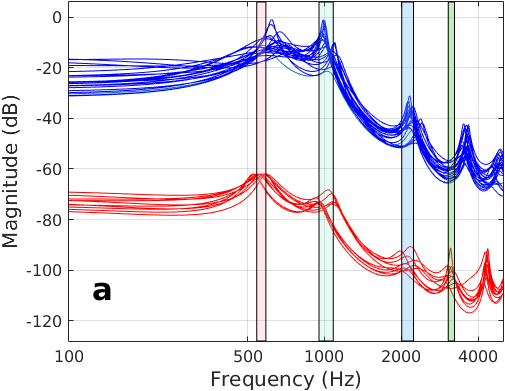}
\includegraphics[width=40mm,height=30mm]{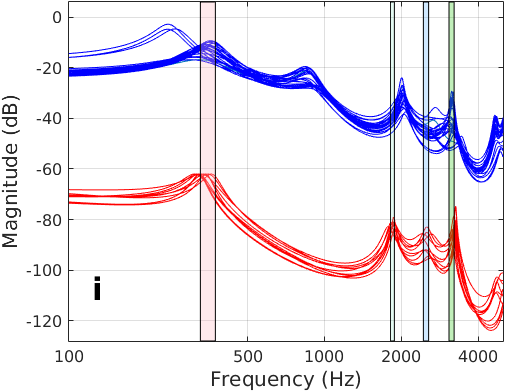}
\end{centering}

\begin{centering}
\includegraphics[width=40mm,height=30mm]{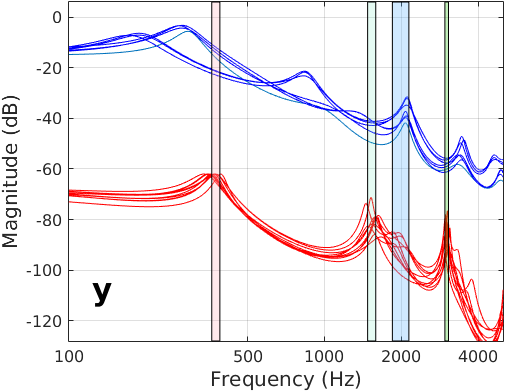}
\includegraphics[width=40mm,height=30mm]{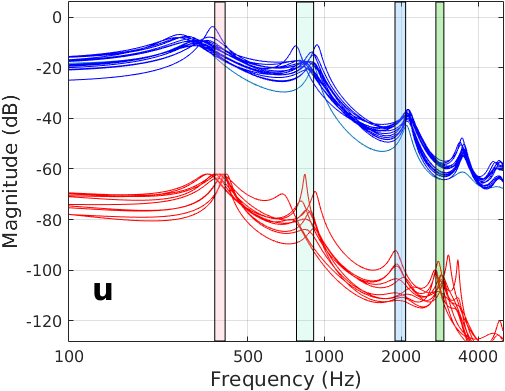}
\end{centering}

\begin{centering}
\includegraphics[width=40mm,height=30mm]{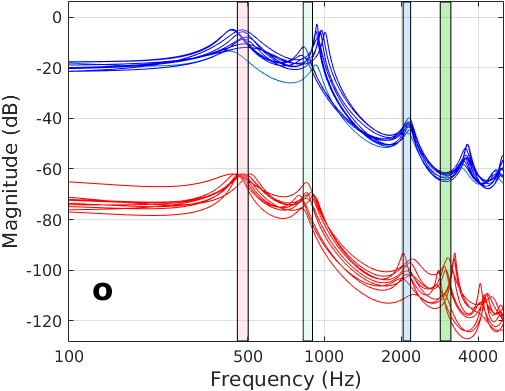}
\includegraphics[width=40mm,height=30mm]{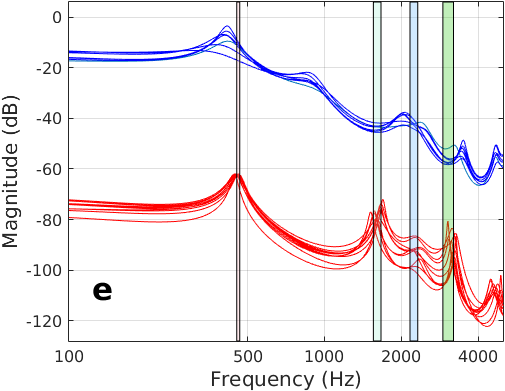}
\end{centering}

\begin{centering}
\includegraphics[width=40mm,height=30mm]{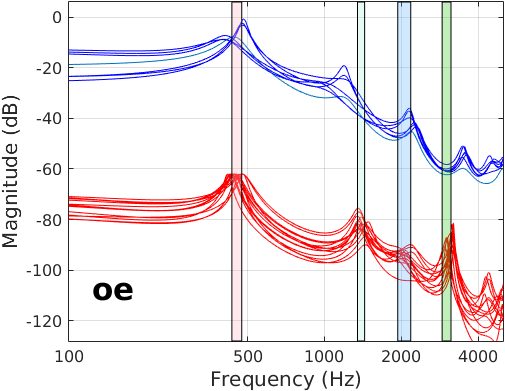}
\includegraphics[width=40mm,height=30mm]{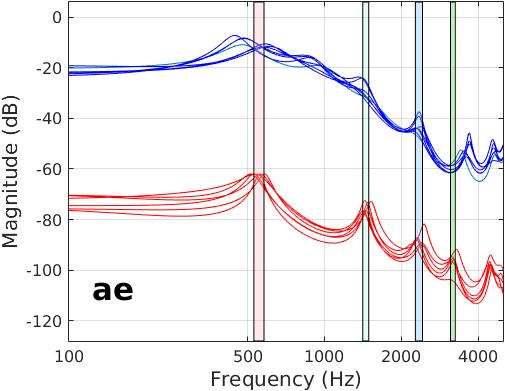}
\end{centering}
\caption{\label{SpectralEnvelopesFront} Spectral envelopes of all
  vowel samples in the dataset. In each panel, the upper curves
  represent post-processed signals recorded during the MRI
  experiments. The lower curves are similar envelopes without any
  post-processing of signals, obtained from the same test subject in
  the anechoic chamber. These two families of curves are comparable to
  curves given in
  \cite[Figs.~7--8]{A-A-H-J-K-K-L-M-M-P-S-V:LSDASMRIS}.  The vertical
  bars are error intervals for formants $F_1, \ldots , F_4$ extracted
  from the recordings in the anechoic chamber.}
\end{figure}

\RevisionText{We extract the vocal tract geometries from the MR images
  by custom software as explained in \cite{AK:AVTMRISM}.  The vocal
  tract geometries are joined with an idealised geometric model of the
  head coil as well as a head geometry as shown in
  Fig.~\ref{exterior}. The head geometry was purchased from
  TurboSquid~\cite{TurboHead}. The computational domain $\Omega$ is
  split into the interior part $\Omega_1$, the exterior part
  $\Omega_2$, and the spherical interface $\Gamma = \partial \Omega_1
  \cap \partial\Omega_2$ as shown in Fig.~\ref{exterior}.  Both
  $\Omega_2$ and $\Gamma$ are same in all computations but $\Omega_1$
  (containing the vowel dependent vocal tract) changes.  }

\RevisionText{We use the finite element method (FEM with piecewise
  linear elements on a tetrahedral mesh with discretisation parameter
  $h >0$) to solve the Helmholtz equation $\Delta u = \kappa^2 u$ in
  $\Omega$ and identify those resonances that have strong excitations
  in $\Omega_2$.  Here $\kappa = \omega / c$ where $c$ is the speed of
  sound, and $\omega$ is the complex angular velocity.
  Using FEM and Nitsche's method (see~\cite{B-H-S:nitsche}) on the interface
  $\Gamma$, the Helmholtz equation takes the variational form }
\begin{equation}
  \label{eq:fem_discretised}
   a(u,v) = \kappa^2 b(u,v) \quad \textrm{for all } v \in V
\end{equation}
\RevisionText{
where the bilinear form $a(\cdot,\cdot)$ is defined as }
\begin{align*}
  a(u,v) &= \sum_{i=1}^2 (\nabla u,\nabla v)_{\Omega_i} - \left<
  \left\{\frac{\partial u}{\partial n}\right\}, \llbracket v
  \rrbracket \right>_\Gamma \\&- \left<\llbracket u \rrbracket,\left\{
  \frac{\partial v}{\partial n} \right\} \right>_\Gamma + \nu_h \left<
  \llbracket u \rrbracket,\llbracket v \rrbracket \right>_\Gamma.
\label{eq:nitsche:a}
\end{align*}
\RevisionText{Here $\{u\}$ ($\llbracket u \rrbracket$) is the average
  (respectively, the jump) of $u$ over the interface $\Gamma$, and
  $\nu_h$ is a mesh size dependent parameter. The bilinear form
  $b(\cdot,\cdot)$ in \eqref{eq:fem_discretised} is the inner product
  of $L^2(\Omega)$. Using Nitsche's method on interface $\Gamma$ makes
  it possible to use the same discretisation of $\Omega_2$ for all
  vowel geometries. For a similar kind of numerical experiment, see \cite{arnela13}.} 

  \RevisionText{The resonance structures of each of the $51$ vowel
    geometries in the data set were computed on $\Omega$ by FEM as
    explained above. The resulting $3060$ complex angular velocities
    $\omega$ were processed as follows:
  \begin{enumerate}
    \item[(i)] Depending on the vowel, three or four $\omega$'s,
      corresponding obviously to the lowest formants of the vocal
      tract volume $\Omega_1$, were excluded.  This was based on
      comparing the energy densities in $\Omega_1$ and $\Omega_2$ of
      the respective eigenfunctions $u$. A total of $2866$ $\omega$'s
      remain that indicate significant acoustic excitation in the
      exterior domain $\Omega_2$.
    \item[(ii)] Next, $1075$ of the $2866$ eigenfunctions $u$ having
      largest $\mathop{Re}{\omega}$ (i.e., being least attenuated)
      were identified, with frequencies between $300 \, \textrm{Hz}
      \ldots 3 \, \textrm{kHz}$.
    \item[(iii)] Eight frequency clusters were formed by the $k$-means
      algorithm (see \cite{M:SMFCAAMO}) from the remaining $1075$
      complex wavenumbers $\omega$ based on the resonant frequencies
      $f = \mathop{Im}{\omega}/2 \pi$.
  \end{enumerate}
The cluster centroids indicate concentrations of acoustic energy
around the eight frequencies, shown by vertical dashed lines in
Fig~\ref{Sweeps}. The energy concentrations coincide quite well with
the peaks of the topmost curve in Fig.~\ref{Sweeps} (last panel),
produced from speech during the MRI. There is much less match with the
middle curve in the same figure, produced from speech in the anechoic
chamber. We conclude that some effects of the MRI coil reflections
are, indeed, present in speech recorded during the MRI.  The
corresponding artefact peaks in speech spectrograms occur at the
frequencies $380 \, \mathrm{Hz}$, $955 \, \mathrm{Hz}$, $1750\,
\mathrm{Hz}$, $2070\, \mathrm{Hz}$, $3230 \, \mathrm{Hz}$, $3970 \,
\mathrm{Hz}$, and $5090 \, \mathrm{Hz}$, of which the four lowest are
displayed as horizontal lines in Fig.~\ref{VowelDiscrepancy}.}

\noindent

\section{Conclusions}

When trying to match a computational model of speech to true speech
biophysics, some sort of paired data is necessary. For example, if the
acoustic modelling is based on vocal tract geometries acquired by MRI,
then the most suitable accompanying data consists of speech samples
recorded during the same MRI scan.  Unfortunately, these samples are
always contaminated by high levels of scanner noise and other acoustic
artefacts that must be eliminated before a reliable extraction of
desired features (such as the formant positions and the spectral tilt)
is possible. Applications related to, e.g., modelling of oral and
maxillofacial surgery require extreme precision that is feasible in
model computations only by careful parameter estimation and validation
of model components. Such models can only be as reliable as their
validation data. 

A post-processing algorithm was proposed for removing acoustic noise
from speech that has been recorded during the MRI using special MRI-proof
instrumentation. It is one of the salient features of MRI scanner
noise that it mainly consists of few strong fundamental frequencies
accompanied by their harmonic overtones. The algorithm outlined in
Section~\ref{CancellationSec} first identifies such harmonic structure
and then adapts a collection of notch filters to the detected
frequencies. The algorithm is realised as MATLAB code.

\RevisionText{ The proposed algorithm is significantly different from
  the approaches presented in
  \cite{bresch,P-H-H:TMMNRRSDPMRID,P-P-F:ASPANPDMRI,I-B-I:TUNSMRIS}.
  Many of these differences are motivated by dissimilarities in
  experimental arrangements for data acquisition.  Scanners with lower
  magnetic field intensity (such as used in
  \cite{P-H-H:TMMNRRSDPMRID,P-P-F:ASPANPDMRI}) typically have an open
  construction where speech may be recorded rather successfully by
  directional microphones, located at a safe distance from the
  scanner. Low-field scanners unfortunately produce worse image
  resolution, and they require longer scanning durations which are
  undesirable features in speech studies.  Here, the recording setup
  is built around a Siemens Magnetom Avanto 1.5T MRI scanner having
  higher magnetic field intensity but a closed construction.  Using
  the arrangement detailed in Fig.~\ref{ArrangementDetail}, we are
  able to obtain an accurate estimate of the scanner noise near the
  test subject's mouth since the MRI coil surfaces act as an
  additional acoustic shield between the speech and the noise
  channels.  Thus, the spectral peaks of noise can be extracted quite
  accurately, and a set of comb filters can be designed to precisely
  and economically remove these frequency bands from speech
  recordings. This makes it unnecessary to resort to methods such as
  the spectral noise gating \cite{I-B-I:TUNSMRIS} or the cepstral
  transformation \cite{P-H-H:TMMNRRSDPMRID} that affect the entire
  frequency range.  Moreover, the proposed algorithm can make good use
  of the fact that our main interest lies in long vowel utterances at
  a fixed $f_0$, chosen not to coincide with the dominant spectral
  peaks of the scanner noise.  The zeroes of the comb filters are
  chosen adaptively for each recording which makes it possible to
  apply the proposed algorithm to different MRI sequences.}

\RevisionText{In our measurement setting, speech and noise samples are
  collected essentially at the same point (see
  Fig.~\ref{ArrangementDetail} and
  \cite{A-A-H-J-K-K-L-M-M-P-S-V:LSDASMRIS}) although from opposite
  directions. Issues related to delays and multiway propagation are
  less serious compared to settings where the sound is collected
  further away as was done in \cite{bresch,P-H-H:TMMNRRSDPMRID}.
  Hence, it is not necessary to develop a high-order noise model as in
  \cite{bresch}, but a computationally less intensive and a more
  tractable post-processing of speech can be used. 
}


\RevisionText{The proposed algorithm operates almost entirely in
  frequency domain which is necessary, regardless of all other
  aspects, for compensating the frequency response of the recording
  system. We point out that also a real-time, time-domain, analogue
  subtraction of MRI noise from recorded speech is used during the
  experiment to provide instant feedback to patient's earphones. The
  analogue circuit removes low frequency noise very effectively but is
  useless at higher frequencies where noise arrives to the sound
  collector channels in different phase. }



The post-processing algorithm was validated by using artificially
noise-contaminated vowels where the noise has been recorded from the
MRI scanner running the same MRI sequence as in the prolonged vowel
experiments. Such artificially MRI noise contaminated vowels have
known formant positions and predetermined SNR's which makes it
possible to assess the achievable noise reduction in post-processing.
In the proposed approach, we observe that $9 \ldots 14 \, \textrm{dB}$
reduction of MRI scanner noise is attainable for prolonged vowel
signals, and the formant extraction error due to post-processing is
less than half a semitone. This is an adequate level of performance
for the validation and the parameter estimation of a computational
speech model such as proposed in \cite{A-A-M-M-V:MLBVFVTOPI}.

The algorithm was applied on real speech data.  A set of prolonged
vowels was recorded during the MRI, and this data was
post-processed. Comparison measurements were recorded in optimal
conditions from the same test subject. Vowel formants were extracted
from both types of data, and it was observed that the formant
discrepancy between the two kinds of data has a strongly frequency
dependent behaviour. Particularly large deviations were observed near
$1 \, \textrm{kHz}$ and $2 \, \textrm{kHz}$. At these frequencies, the
formant discrepancy is several times as large as the formant
estimation error due to the post-processing algorithm, and the
deviations are statistically significant (Student's t-test with $p >
0.95$). We presented computational evidence that the deviant
frequencies are related to the acoustic resonances of the space
between test subject's face and MRI coils.  However, some of the
formant error may also be due to test subject's adaptation to his
acoustic environment during the MRI scan.

The notch filtering adds a large number of transmission zeros to
processed signals which causes the phase response of the algorithm to
be non-linear. This may be a showstopper if the post-processed signal
is to be used as an input for another speech processing algorithm such
as the Glottal Inverse Filtering (GIF) for glottal pulse extraction,
see \cite{Alku:2011:IFreview,ALKU:IAIF}.  To produce signals with
linear phase response, one should use, e.g., non-causal spectral
filtering (see \cite{G-R:NAPMVS}) instead of notch filters.

Even though the algorithm has been designed for the main purpose of
formant extraction, it gives audibly quite satisfactory results from
natural speech that has been recorded during dynamic MRI of
mid-sagittal sections.

\section*{Acknowledgements}

 The authors wish to thank many collegues for consultation and
 facilities: Dept. Signal Processing and Acoustics, Aalto University
 (Prof.~P.~Alku), PUMA research group at Dept. Oral and Maxillofacial
 Surgery, University of Turku (Prof.~R.-P.~Happonen and Dr.~D.~Aalto),
 Medical Imaging Centre of Southwest Finland (Prof.~R.~Parkkola and
 Dr.~J.~Saunavaara), and Aalto University Digital Design Laboratory
 (Mr.~A.~Mohite). \RevisionText{The authors wish to
   express their gratitude to the three anonymous reviewers
 for their comments and ideas for improvements.}

 The authors have received financial support from Instrumentarium
 Science Foundation, Vilho, Yrj\"o and Kalle V\"ais\"al\"a Foundation,
 and Magnus Ehrnrooth Foundation. 



\end{document}